\documentclass[final,5p,times,twocolumn]{elsarticle}

\usepackage{amssymb}
\usepackage{amsmath}
\usepackage{graphicx}
\usepackage{subcaption}
\usepackage{booktabs}
\usepackage{array}
\usepackage[table]{xcolor}
\usepackage{colortbl}
\usepackage{siunitx}   
\usepackage{float}
\usepackage{placeins}      

\usepackage{dblfloatfix}   
\usepackage[hidelinks]{hyperref}

\journal{SAE Technical Paper 20XX-01-XXXX}

\begin{document}

\begin{frontmatter}

\title{CFD-Machine Learning Driven Pod Optimization and Staged Pressure--Area Management for Supersonic Evacuated Tube Transport}

\author{Jaydip Patel}
\author{Dhwanil Shukla}
\affiliation{organization={Indian Institute of Technology Bombay},
            addressline={Powai},
            city={Mumbai},
            postcode={400076},
            state={Maharashtra},
            country={India}}

\begin{abstract}
This study investigates the aerodynamic feasibility of high-speed evacuated tube transport (ETT) using an integrated CFD--machine learning framework for pod geometry optimization and a staged converging--diverging (CD) tube concept for supersonic operation. The pod geometry is parameterized using composite cubic B\'{e}zier curves, with seven design variables controlling the nose, body, and tail profiles. Latin Hypercube Sampling is used to generate a uniformly distributed design space across subsonic and supersonic operating regimes. Two-dimensional axisymmetric ANSYS Fluent simulations, validated against published compressible-flow data with an approximately 1\% discrepancy at the primary validation speed, are used to evaluate the aerodynamic drag of 200 pod configurations. The resulting CFD dataset is used to train surrogate models for rapid drag prediction, among which XGBoost provides the best overall performance, achieving a test $R^{2}$ of 0.9524 with the lowest generalization gap among the benchmarked regressors. The trained surrogate is then coupled with Differential Evolution and Particle Swarm Optimization to identify low-drag pod geometries, which are subsequently re-evaluated using CFD to confirm the surrogate predictions.

Beyond pod-shape optimization, the study examines a staged CD tube architecture that combines controlled area variation with gradual pressure reduction to mitigate choking during acceleration to supersonic speeds. In the idealized axisymmetric simulations considered here, the proposed strategy enables transition from atmospheric entry at Mach~1.6 to near-vacuum cruise at Mach~2.5 and blockage ratio $\beta = 0.50$ without the formation of a choked annular flow region. The simulations show a reduction in total drag from 68{,}913.7\,N under atmospheric supersonic conditions to 120.5\,N in the final near-vacuum cruise stage, primarily due to the transition from wave-drag-dominated resistance to viscous shear-dominated resistance. These results indicate that combined pod-shape optimization and staged pressure--area management can substantially reduce aerodynamic resistance in supersonic ETT systems. The findings provide a concept-level aerodynamic basis for further investigation of supersonic Hyperloop operation, including three-dimensional effects, transient acceleration dynamics, structural constraints, propulsion requirements, and full-system energy costs.
\end{abstract}


\end{frontmatter}

\section{Introduction}
\label{sec:introduction}

Transportation systems are essential to modern economies, but they remain a major source of energy consumption and greenhouse gas emissions \cite{IEA2021}. A persistent challenge in high-speed transport is that increasing travel speed generally leads to disproportionately higher aerodynamic drag and propulsion power requirements. This makes it difficult to simultaneously achieve high speed, energy efficiency, and environmental sustainability.

Evacuated Tube Transport (ETT), commonly referred to as Hyperloop, has been proposed as a potential solution to this challenge. The concept, introduced in the Hyperloop Alpha proposal \cite{Musk2013}, involves transporting pods through partially evacuated tubes at very low ambient pressure, thereby reducing aerodynamic drag. In principle, this approach can combine the speed advantages of aviation with the energy-efficiency benefits of rail, provided the associated aerodynamic, structural, and operational challenges are addressed.

Although ETT is often discussed for high-speed subsonic travel, its strongest value proposition may lie in enabling supersonic ground transport. Conventional supersonic civil aviation remains limited by high propulsion power, emissions, and noise constraints, particularly sonic boom. In contrast, the low-pressure confined environment of ETT systems modifies the aerodynamic constraints and may offer a more feasible pathway for sustained supersonic operation. If ETT remains limited to subsonic speeds, its advantage over advanced rail and aviation systems becomes less distinct.

However, supersonic ETT operation introduces fundamental aerodynamic challenges. The pod moves through a confined annular passage between the pod surface and the tube wall, making the flow strongly dependent on pod geometry, operating Mach number, tube pressure, and blockage ratio. As blockage ratio increases, the available annular flow area decreases, promoting local acceleration, compression waves, shock formation, and eventually flow choking. Once choking occurs, upstream pressure buildup and rapid drag increase can severely limit further acceleration \cite{Lang2024}. Mitigating this choking constraint is therefore essential for efficient supersonic ETT operation.

The present study addresses this problem through two coupled objectives. First, it develops a CFD--machine learning framework for systematic aerodynamic optimization of Hyperloop pod geometries across subsonic and supersonic operating regimes. Second, it examines a staged converging--diverging (CD) tube concept as a potential system-level approach for mitigating choking during acceleration to supersonic speeds. The framework integrates two-dimensional axisymmetric CFD simulations, surrogate modeling, and global optimization algorithms to identify low-drag pod geometries and assess staged pressure--area management under idealized operating conditions.

This work is intended as a concept-level aerodynamic study rather than a complete system-level design of a deployable ETT network. The analysis focuses on aerodynamic trends associated with pod shape, confinement, pressure level, and Mach number. The results provide a basis for further investigations involving three-dimensional effects, transient acceleration dynamics, propulsion integration, structural constraints, airlock dynamics, and full-system energy assessment.

\subsection{Related Work}
\label{sec:background}

The aerodynamic environment of Hyperloop systems differs significantly from conventional external vehicle aerodynamics because of the combined effects of low ambient pressure, high speed, and geometric confinement. These conditions lead to strong compressibility effects, shock interactions, boundary-layer development, and choking in the annular passage between the pod and tube wall.

Previous studies have identified blockage ratio, pod velocity, and tube pressure as the dominant parameters controlling aerodynamic performance. Increasing blockage ratio reduces the critical Mach number and causes earlier onset of choking, leading to rapid growth in pressure drag \cite{Oh2019}. Analytical and numerical studies have further quantified the relationship between blockage ratio and critical Mach number, identifying shock formation and drag rise near the choking limit \cite{Zhou2020}. Detailed simulations have shown that compression and expansion waves produce distinct flow regimes, including pre-choking, choking, and shock-dominated operation \cite{Le2020,Jang2021}. These constraints are often interpreted through the Kantrowitz limit, which restricts feasible operating conditions in confined high-speed flows \cite{Zhou2022}.

Aerodynamic shape optimization has therefore been widely explored as a route to drag reduction. Early studies using simplified two-dimensional geometries showed that streamlined nose and tail profiles can significantly reduce drag \cite{Chen2012,Gillani2019}. More recent three-dimensional studies demonstrated that tail geometry can strongly influence total drag, while head and tail effects can often be considered with partial independence \cite{Le2022}. Studies including transition modeling have also highlighted the importance of boundary-layer behavior and the trade-off between pressure drag and skin-friction drag \cite{Nick2020}.

System-level choking mitigation strategies have also been proposed. One prominent approach is the use of onboard compressors to transfer air from the front of the pod to the rear, thereby reducing upstream pressure buildup and extending the operating range. While effective in principle, onboard compression introduces mechanical complexity, additional power requirements, and integration challenges \cite{Bizzozero2021}.

Despite these advances, two limitations remain important. First, many aerodynamic optimization studies rely on repeated high-fidelity CFD simulations, which limits the extent of design-space exploration. Surrogate modeling can reduce this cost, but its integration with systematic pod-shape optimization across both subsonic and supersonic regimes remains limited. Second, while choking is widely recognized as a major barrier to high-speed ETT operation, practical alternatives to onboard compression have received comparatively less attention.

The present study addresses these gaps by combining CFD-based aerodynamic analysis, machine-learning surrogate modeling, and population-based global optimization for pod design. In parallel, it introduces a staged CD tube concept in which tube area variation and pressure reduction are coordinated to manage compressible flow around the pod. This approach is evaluated as a potential strategy to reduce choking risk during transition from atmospheric entry to near-vacuum supersonic cruise, without relying on onboard compressor systems.

\subsection{Scope of the Present Work}

This study proposes a hybrid CFD--machine learning framework for aerodynamic drag reduction in Hyperloop pod design. Two-dimensional axisymmetric simulations are performed in ANSYS Fluent to generate a structured dataset over a systematically varied design space. The pod profile is parameterized using composite cubic B\'{e}zier curves, with geometric design variables controlling the nose, body, and tail shapes. The resulting CFD-generated drag data are used to train machine-learning surrogate models for rapid aerodynamic prediction.

The trained surrogate model is coupled with global optimization algorithms to identify low-drag pod geometries for both subsonic and supersonic operating regimes. The optimized designs are subsequently re-evaluated using CFD to assess agreement between surrogate predictions and full-order simulations.

In addition to pod-shape optimization, the study introduces a converging--diverging tube concept for supersonic ETT operation. The proposed approach uses staged tube-area variation and controlled pressure reduction to manage flow acceleration and reduce the likelihood of choking in the annular region around the pod. The concept is inspired by area variation in compressible internal flows and is evaluated here within an idealized axisymmetric CFD framework. Overall, the study provides a computationally efficient pod-optimization workflow and a preliminary aerodynamic assessment of staged pressure--area management for supersonic ETT systems.

\section{Theoretical Framework}
\label{sec:theoretical_framework}

The aerodynamic behaviour of a Hyperloop pod moving through a partially evacuated tube is governed primarily by the interaction between pod speed, tube confinement, and gas compressibility. Two parameters are especially important: the blockage ratio $\beta$, which measures the degree of geometric confinement, and the pod Mach number $M_\text{pod}$, which determines the strength of compressibility effects. Together, these parameters define the feasible operating envelope of the system and determine whether the annular flow around the pod remains unchoked or transitions to a choked, shock-dominated state. This section summarizes the theoretical basis used to interpret the CFD results and to motivate the proposed staged converging--diverging tube concept.

\subsection{Blockage Ratio and the Piston Effect}

In the pod-fixed reference frame, the surrounding air moves toward the pod and is forced through the annular passage between the pod surface and the tube wall. The pod therefore acts as a moving blockage inside a confined duct. This produces a piston effect, characterized by pressure rise ahead of the pod, pressure reduction in the wake region, and an associated pressure-drag penalty. The severity of this confinement is quantified using the blockage ratio,

\begin{equation}
    \beta = \frac{A_\text{pod}}{A_\text{tube}},
    \label{eq:blockage}
\end{equation}

\noindent where $A_\text{pod}$ is the maximum frontal cross-sectional area of the pod and $A_\text{tube}$ is the internal cross-sectional area of the tube. A larger blockage ratio reduces the available annular flow area and increases the acceleration of air around the pod. Under subsonic and transonic conditions, this generally increases both pressure drag and the risk of choking. Conversely, a smaller blockage ratio is aerodynamically favourable, but requires either a smaller pod cross-section or a larger tube diameter, both of which have practical implications for passenger capacity and infrastructure cost.

In the near-vacuum supersonic regime considered later in this study, the relation between blockage ratio and drag becomes more nuanced. Since the density is substantially reduced, the absolute aerodynamic resistance is much lower than under atmospheric conditions. As discussed in Section~\ref{sec:blockage_sweep}, the simulations indicate that increasing $\beta$ in the final near-vacuum cruise regime can reduce the drag coefficient by modifying the annular acceleration and pressure recovery characteristics. This behaviour should be interpreted within the specific low-pressure operating conditions considered here, rather than as a general result for all ETT regimes.

\subsection{Validity of the Continuum Assumption}

The proposed operating conditions involve tube pressures as low as approximately 100~Pa. At such pressures, the gas density is much lower than at standard atmospheric conditions, and it is therefore necessary to verify that the continuum approximation remains valid. The relevant nondimensional parameter is the Knudsen number, defined as the ratio of the molecular mean free path $\lambda$ to a representative macroscopic length scale $h$:

\begin{equation}
    Kn = \frac{\lambda}{h}
    = \frac{k_B T}{\sqrt{2}\,\pi d^2\, h\, p},
    \label{eq:knudsen}
\end{equation}

\noindent where $k_B$ is the Boltzmann constant, $T$ is the gas temperature, $d$ is the effective molecular diameter, and $p$ is the static pressure. Using representative values of $h = 2$~m, $T = 300$~K, $d = 3.7 \times 10^{-10}$~m, and $p = 100$~Pa gives $Kn \sim 10^{-5}$. This value is well below the commonly used continuum threshold of $Kn < 10^{-2}$. Therefore, the Navier--Stokes equations are considered appropriate for the operating pressures and length scales examined in this work.

\subsection{One-Dimensional Compressible Flow Model}

The confined flow around the pod can be interpreted, in an idealized one-dimensional sense, as flow through a variable-area duct. In the pod-fixed frame, the annular area $A(x)$ decreases as the flow approaches the maximum pod cross-section and then increases downstream. This behaviour is analogous to a converging--diverging passage, with the minimum annular gap acting as an effective throat. Although the actual flow is viscous, multidimensional, and affected by boundary layers and shocks, the one-dimensional inviscid model provides useful estimates of choking limits and operating regimes.

Under the assumptions of steady, inviscid, adiabatic, and isentropic flow, the ideal-gas and stagnation relations are given by

\begin{align}
    p &= \rho R T, 
    \label{eq:ideal_gas}\\[4pt]
    a &= \sqrt{\gamma R T}, 
    \label{eq:sos}\\[4pt]
    \frac{T_0}{T} &= 1 + \frac{\gamma-1}{2}M^2,
    \label{eq:total_temp}\\[4pt]
    \frac{p_0}{p} &= 
    \left(1 + \frac{\gamma-1}{2}M^2
    \right)^{\frac{\gamma}{\gamma-1}},
    \label{eq:total_pres}
\end{align}

\noindent where $p$, $\rho$, and $T$ are the local static pressure, density, and temperature, respectively; $R$ is the specific gas constant; $\gamma$ is the ratio of specific heats; $a$ is the local speed of sound; $M$ is the local Mach number; and the subscript $0$ denotes stagnation quantities.

The corresponding mass flow rate through a cross-sectional area $A$ can be written as

\begin{equation}
    \dot{m} = M A p_0 
    \sqrt{\frac{\gamma}{R T_0}}
    \left(1 + \frac{\gamma-1}{2}M^2
    \right)^{-\frac{\gamma+1}{2(\gamma-1)}}.
    \label{eq:massflow}
\end{equation}

This expression shows that, for fixed stagnation conditions and area, the mass flux reaches a maximum at $M=1$. Once sonic conditions occur at the minimum annular area, the flow becomes choked and the mass flow rate cannot increase further without changes in upstream stagnation conditions or effective flow area.

The local relationship between area change and velocity change is described by the area--velocity relation,

\begin{equation}
    \frac{dA}{A} = \left(M^2 - 1\right)\frac{du}{u},
    \label{eq:area_vel}
\end{equation}

\noindent where $u$ is the local flow velocity. For subsonic flow ($M<1$), a decrease in area accelerates the flow, whereas an increase in area decelerates it. For supersonic flow ($M>1$), the opposite behaviour occurs. Thus, a continuous isentropic transition from subsonic to supersonic flow requires sonic conditions at the minimum-area location. In the present ETT context, this minimum-area location corresponds approximately to the narrowest annular gap around the pod.

\subsection{Critical Limits and Operating Regimes}

The one-dimensional model provides two important theoretical limits in the $(\beta,\,M_\text{pod})$ space: the isentropic limit and the Kantrowitz limit. These limits are not substitutes for CFD analysis, but they provide a useful framework for interpreting when the annular flow is expected to remain unchoked, become choked, or admit multiple possible steady states depending on the operating path.

\subsubsection{Isentropic Limit}

The isentropic limit corresponds to the onset of choking in an ideal, shock-free flow. Choking occurs when the local Mach number reaches unity at the minimum annular area. Applying the area--Mach relation between the upstream flow condition, where the Mach number is associated with the pod speed $M_\text{pod}$, and a sonic throat condition $M^*=1$, gives the blockage-ratio limit

\begin{equation}
    \beta_\text{isen} = 1 - M_\text{pod}
    \left(1 + \frac{\gamma-1}{2}M_\text{pod}^2
    \right)^{-\frac{\gamma+1}{2(\gamma-1)}}
    \left(\frac{\gamma+1}{2}\right)^{\frac{\gamma+1}{2(\gamma-1)}}.
    \label{eq:isentropic_limit}
\end{equation}

For a given pod Mach number, blockage ratios below this limit allow an unchoked isentropic solution. When this limit is exceeded, the annular throat reaches sonic conditions, the mass flow rate past the pod becomes constrained, and upstream pressure buildup can occur. In practical ETT operation, crossing this limit is associated with a rapid rise in aerodynamic drag and the possible formation of shock structures ahead of or around the pod.

\subsubsection{Kantrowitz Limit}

At sufficiently high pod speeds, the flow may transition from a choked state to a started supersonic state, in a manner analogous to the starting process of a supersonic inlet. This behaviour is represented by the Kantrowitz limit. The limit is obtained by considering a normal shock upstream of the converging section and imposing a choked-throat condition downstream, while accounting for the total-pressure loss across the shock \cite{Lang2024}. The corresponding blockage-ratio limit is

\begin{equation}
    \beta_\text{Kant} = 1 -
    \left[\frac{(\gamma+1)M_\text{pod}^2}
    {(\gamma-1)M_\text{pod}^2+2}\right]^{-\frac{1}{2}}
    \left[\frac{(\gamma+1)M_\text{pod}^2}
    {2\gamma M_\text{pod}^2-(\gamma-1)}
    \right]^{-\frac{1}{\gamma-1}}.
    \label{eq:kantrowitz_limit}
\end{equation}

For blockage ratios below the Kantrowitz limit, a fully started supersonic solution is theoretically possible. However, this regime generally requires very low blockage ratios, which may be impractical for passenger-scale ETT systems because it would require either a very small pod or a very large tube diameter. For practically meaningful pod sizes, the system may instead operate in a region where both choked and unchoked solutions are theoretically admissible.

\subsubsection{Dual-Solution Zone and Path Dependence}

\begin{figure}[htbp]
    \centering
    \includegraphics[width=\columnwidth]{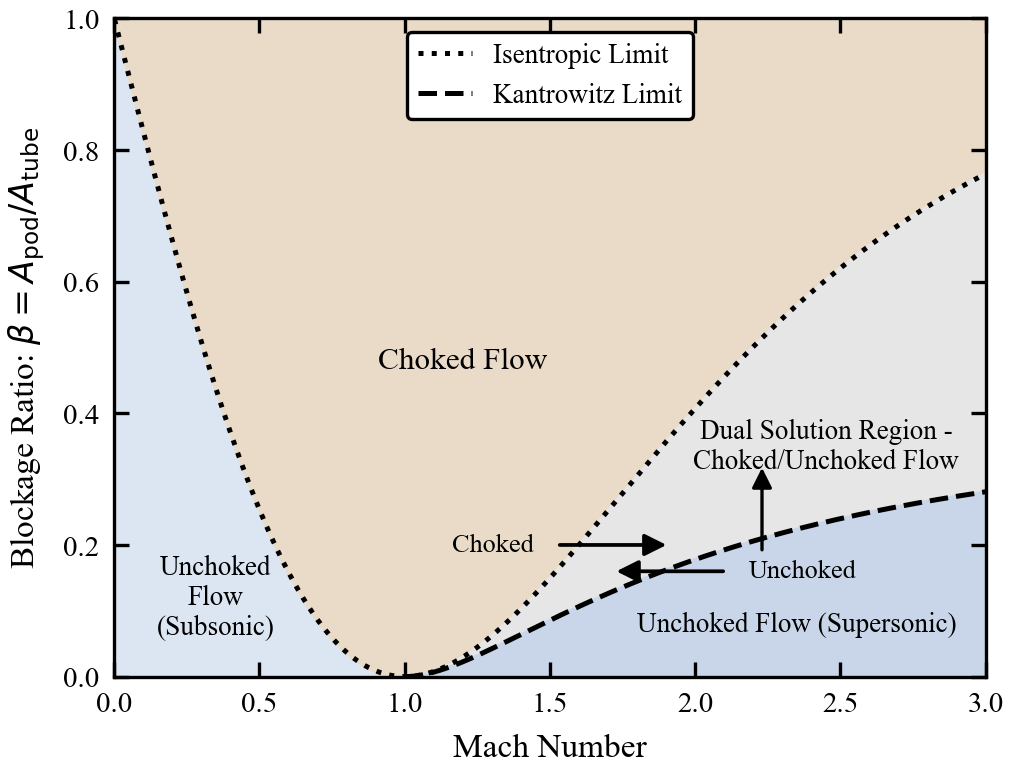}
    \caption{Flow-regime map for $\gamma = 1.4$ constructed using the isentropic limit in Eq.~\ref{eq:isentropic_limit} and the Kantrowitz limit in Eq.~\ref{eq:kantrowitz_limit}. The region between the two limits represents the dual-solution zone, where the realized flow state may depend on the operating path.}
    \label{fig:regime_map}
\end{figure}

The isentropic and Kantrowitz limits together define distinct regions in the $(\beta,\,M_\text{pod})$ operating map, as shown in Figure~\ref{fig:regime_map}. The region between these two limits is referred to here as the dual-solution zone. In this region, two different steady flow states may be possible for the same nominal operating point: an unchoked, shock-free state and a choked state involving shock-induced total-pressure loss.

The actual state realized depends on the operating history. If the pod enters the dual-solution zone from a low-speed or low-blockage unchoked condition and the isentropic limit is not crossed, the flow can remain on the unchoked branch. However, if the operating path crosses the isentropic limit, the annular throat becomes sonic and the flow transitions to a choked state. Once this transition occurs, returning to an unchoked state may require a different operating path, similar to the hysteresis associated with supersonic inlet starting.

This path dependence motivates the staged operating strategy considered in the present study. Instead of attempting to operate directly at high Mach number and high blockage ratio, the system can first accelerate along a lower-blockage trajectory and then increase the effective blockage ratio while remaining within the unchoked branch of the dual-solution zone. In the proposed CD tube concept, this is interpreted as a coordinated variation of tube area and pressure level during acceleration from atmospheric entry to near-vacuum supersonic cruise. The isentropic limit therefore acts as the critical boundary for avoiding choking, while the Kantrowitz limit helps identify the range over which multiple flow states may be possible.

\section{Methodology}
\label{sec:methodology}

The methodology adopted in this study integrates parametric geometry generation, CFD-based aerodynamic evaluation, machine-learning surrogate modelling, and population-based global optimisation. The overall workflow is designed to identify low-drag Hyperloop pod geometries for both subsonic and supersonic operating regimes while reducing the computational cost associated with repeated high-fidelity CFD simulations.

The framework consists of five main stages. First, the pod dimensions are constrained using passenger-capacity and interior-layout requirements. Second, smooth axisymmetric pod profiles are generated using a composite cubic B\'{e}zier parameterisation. Third, the design space is sampled using Latin Hypercube Sampling (LHS), and each sampled geometry is evaluated using two-dimensional axisymmetric CFD simulations in ANSYS Fluent. Fourth, the CFD-generated dataset is used to train and benchmark surrogate regression models for predicting aerodynamic drag from geometric parameters. Finally, the best-performing surrogate is coupled with Differential Evolution (DE) and Particle Swarm Optimisation (PSO) to search for minimum-drag pod geometries. The final optimised designs are reconstructed and re-evaluated using CFD to verify the surrogate-based predictions.

This workflow preserves the physical fidelity of CFD for data generation and final validation, while using machine learning to accelerate the design-space exploration and optimisation stages.

\subsection{Pod Geometry and Design Space}
\label{sec:pod_geometry}

\subsubsection{Interior Sizing and Physical Constraints}
\label{Intsizing}

The starting point for pod design is the passenger-capacity requirement. In this study, the pod is sized to accommodate approximately 20--25 passengers while maintaining a compact aerodynamic envelope suitable for confined-tube operation. Geometric references are drawn from the Airbus A320 narrow-body cabin, which provides a practical benchmark for single-aisle passenger layout and seating dimensions.

A 1+1 seating arrangement, with one seat on each side of a central aisle, is adopted to determine the minimum usable cabin width. The required width accounts for seat width, armrest clearance, aisle clearance, and structural wall thickness. The vertical envelope is determined by considering the stacked contributions of floor structure thickness, magnetic-levitation clearance beneath the floor, seated and standing passenger height requirements, overhead clearance, upper duct allowance, and structural shell thickness.

The resulting internal height requirement exceeds the internal width requirement. Therefore, the axisymmetric body-of-revolution geometry is selected by allowing the larger dimension, namely the required vertical height, to determine the pod radius. This choice satisfies both width and height constraints while avoiding the additional aerodynamic and structural complexity of a non-circular cross-section.

Based on these considerations, the following geometric constraints are imposed throughout the design process:

\begin{itemize}
    \item Maximum pod radius: $H_{\max} = 1.2$\,m, corresponding to a pod diameter of 2.4\,m.
    \item Body length: $l_b \in [12.0,\;16.0]$\,m, ensuring sufficient cabin length for the target passenger capacity.
\end{itemize}

\noindent
Since the maximum pod radius is fixed, the frontal cross-sectional area remains constant across all generated designs. Aerodynamic optimisation is therefore achieved through variations in the longitudinal profile, rather than by reducing the passenger-carrying cross-section.

\subsubsection{Composite B\'{e}zier Parameterisation}
\label{sec:bezier_parameterisation}

The pod profile is decomposed into three aerodynamically distinct regions: nose, cylindrical body, and tail. Each region is represented using a cubic B\'{e}zier curve segment. The complete profile is formed by joining these segments into a composite B\'{e}zier curve with first-derivative continuity at the nose--body and body--tail interfaces. This $C^1$ continuity ensures that the profile slope is continuous across segment junctions, avoiding abrupt geometric changes that could introduce nonphysical pressure gradients, mesh-quality issues, or numerical artefacts in the CFD solution.

The upper meridional profile generated from the B\'{e}zier construction is revolved about the longitudinal axis to define the corresponding axisymmetric pod surface. For CFD preprocessing, each geometry is exported as a discrete point-cloud representation of the meridional profile and used to construct the two-dimensional axisymmetric flow domain.

\begin{figure}[htbp]
    \centering
    \includegraphics[width=0.8\columnwidth]{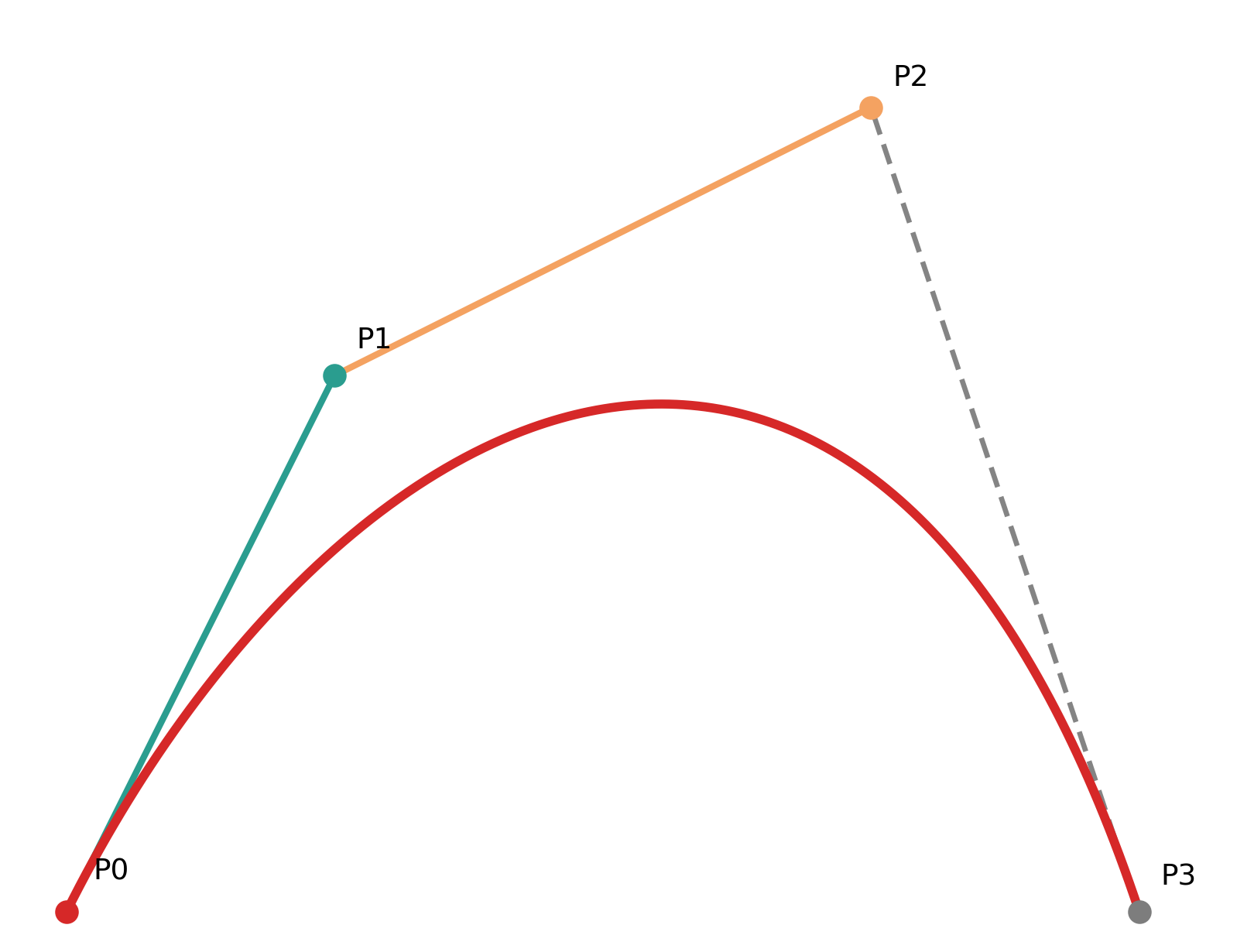}
    \caption{Composite cubic B\'{e}zier representation used for smooth pod-profile generation.}
    \label{fig:bezier}
\end{figure}

\subsubsection{Design Variables and Bounds}
\label{sec:design_variables}

Each pod profile is defined by seven geometric design variables: nose length, body length, tail length, junction radius, shoulder position, nose angle, and tail angle. These parameters provide physically interpretable control over the main aerodynamic features of the pod, including forebody slenderness, axial location of maximum radius, aft-body closure rate, and local profile smoothness near segment junctions.

The design bounds are selected separately for the subsonic and supersonic regimes because the dominant aerodynamic constraints differ between these regimes. In the subsonic case, the geometry must reduce pressure drag while avoiding excessive wetted area. In the supersonic case, the nose and tail must additionally control shock formation, wave drag, and shock-induced separation. The design-variable bounds used in the present study are summarised in Table~\ref{tab:design_bounds}.

\begin{table}[htbp]
\centering
\small
\renewcommand{\arraystretch}{1.25}
\setlength{\tabcolsep}{6pt}
\begin{tabular}{lcc}
\hline
\textbf{Design variable} & \textbf{Subsonic regime} & \textbf{Supersonic regime} \\
\hline
Nose length, $l_\text{nose}$       & $2.0 - 5.0$\,m     & $5.0 - 10.0$\,m  \\
Body length, $l_\text{body}$       & $12.0 - 16.0$\,m   & $12.0 - 16.0$\,m \\
Tail length, $l_\text{tail}$       & $3.5 - 7.0$\,m     & $6.0 - 12.0$\,m  \\
Junction radius, $h_j$             & $0.85 - 0.95$\,m   & $0.85 - 0.95$\,m \\
Shoulder position, $s_\text{pos}$  & $0.15 - 0.70$      & $0.35 - 0.65$    \\
Nose angle, $\alpha_\text{nose}$   & $19^{\circ} - 44^{\circ}$ & $10^{\circ} - 21^{\circ}$ \\
Tail angle, $\alpha_\text{tail}$   & $14^{\circ} - 29^{\circ}$ & $8^{\circ} - 18^{\circ}$  \\
\hline
\end{tabular}
\caption{Design-variable bounds used for subsonic and supersonic pod-geometry
generation.}
\label{tab:design_bounds}
\end{table}

The body length and maximum pod radius are constrained by passenger-layout requirements and are not relaxed solely for aerodynamic convenience. Nose and tail lengths are allowed to vary over wider ranges because they directly influence pressure recovery, boundary-layer development, and shock structure. In the subsonic regime, the design space includes a range of rounded, blunt, and streamlined forebody shapes to provide topological diversity in the training data. In the supersonic regime, longer and sharper forebodies are favoured to promote attached oblique shock formation and reduce wave drag, while longer tail sections are used to limit adverse pressure gradients and reduce the likelihood of shock-induced separation in the annular passage.

For the supersonic case, an additional nose-tip control factor is used to regulate the sharpness of the leading-edge geometry. This factor is implemented as a normalised B\'{e}zier control parameter varying between 0.05 and 0.20. Lower values generate a sharper, near-tangent nose profile resembling a von K\'{a}rm\'{a}n-type ogive, while higher values produce a more rounded forebody. This parameter allows the optimiser to adjust the nose-tip sharpness without prescribing a fixed shock angle a priori.Nose and tail angles are not sampled directly but are derived from
the junction radius $h_j$ and respective length bounds via
$\alpha = \arctan\!\bigl(h_j \,/\, (l/2)\bigr)$; the ranges shown reflect the
extremes implied by the sampled parameter space.

\subsubsection{Design-Space Sampling}
\label{sec:lhs_sampling}

The design space is sampled using Latin Hypercube Sampling (LHS). LHS is a stratified space-filling method in which each design variable is divided into equal-probability intervals and one sample is drawn from each interval. Compared with simple random sampling, LHS provides more uniform coverage of the design space for the same number of samples and reduces clustering in high-dimensional parameter spaces.

In the present study, LHS is used to generate geometrically distinct pod profiles for both subsonic and supersonic operating regimes. This ensures that the surrogate model is trained on a broad range of feasible shapes and is not biased toward a small region of the design space. The space-filling nature of LHS is particularly important because the surrogate model is later used inside an optimisation loop, where reliable interpolation across the feasible design domain is required.

\subsection{CFD Simulation Setup}
\label{sec:cfd_setup}

\subsubsection{Computational Domain and Reference Frame}
\label{sec:domain_reference_frame}

Aerodynamic evaluation of each pod geometry is performed using two-dimensional axisymmetric CFD simulations in ANSYS Fluent. The axisymmetric formulation is appropriate because the generated pod geometries are bodies of revolution. This reduces the computational problem to the meridional plane while retaining the dominant effects of pod shape, annular confinement, compressibility, and near-wall viscous behaviour.

A pod-fixed reference frame is used. In this frame, the pod remains stationary and the surrounding gas moves past it at the prescribed operating velocity. The tube wall is assigned a moving-wall boundary condition with velocity equal to the freestream speed. This treatment is equivalent to pod motion through a stationary tube and avoids the need for dynamic or overset mesh methods during the large number of simulations required for dataset generation.

The computational domain extends 60\,m upstream of the pod nose and 100\,m downstream of the pod tail. These lengths are chosen to minimise the influence of inlet and outlet boundaries on the flow around the pod and to allow wake structures and pressure disturbances to develop sufficiently before reaching the outlet. The blockage ratio is held fixed within each optimisation campaign: $\beta = 0.36$ for the subsonic regime and $\beta = 0.20$ for the supersonic regime. The tube diameter is adjusted accordingly to maintain the prescribed blockage ratio for the fixed maximum pod radius.

\subsubsection{Boundary Conditions and Physical Models}
\label{sec:boundary_conditions}

The boundary conditions used in the CFD simulations are summarised in Table~\ref{tab:bc}. The inlet/far-field condition imposes the prescribed operating Mach number and static pressure. The outlet pressure is set consistently with the far-field static pressure. The pod surface is treated as a stationary no-slip adiabatic wall, while the tube wall is treated as a moving no-slip wall in the pod-fixed frame. The lower boundary of the meridional domain is assigned an axis condition.

\begin{table}[htbp]
\centering
\small
\renewcommand{\arraystretch}{1.15}
\setlength{\tabcolsep}{4pt}
\begin{tabular}{ll}
\hline
\textbf{Boundary} & \textbf{Specification} \\
\hline
Inlet / far-field boundary & Prescribed static pressure and Mach number \\
Tube wall                  & Moving no-slip wall\\
Pod surface                & Stationary no-slip adiabatic wall \\
Outlet                     & Prescribed static pressure \\
Axis                       & Axisymmetric boundary condition \\
\hline
\end{tabular}
\caption{Boundary conditions used in the two-dimensional axisymmetric CFD simulations.}
\label{tab:bc}
\end{table}

The gas is modelled as an ideal gas, and dynamic viscosity is evaluated using Sutherland's law. The simulations are performed at a static temperature of 300\,K with air properties defined using $R=287$\,J/kg$\cdot$K and $\gamma = 1.4$. The main flow and physical model parameters are listed in Table~\ref{tab:flow_parameters}.

\begin{table}[htbp]
\centering
\small
\renewcommand{\arraystretch}{1.25}
\setlength{\tabcolsep}{6pt}
\begin{tabular}{ll}
\hline
\textbf{Parameter} & \textbf{Value / model} \\
\hline
Freestream velocity     & Regime dependent \\
Static pressure         & 101.325\,Pa \\
Temperature             & 300\,K \\
Gas constant, $R$       & 287\,J/kg$\cdot$K \\
Ratio of specific heats, $\gamma$ & 1.4 \\
Density                 & Ideal-gas law \\
Dynamic viscosity       & Sutherland's law \\
Turbulence model        & SST $k$--$\omega$ \\
\hline
\end{tabular}
\caption{Flow conditions and physical models used in the CFD simulations.}
\label{tab:flow_parameters}
\end{table}

\begin{figure*}[htbp]
    \centering
    \includegraphics[width=\textwidth]{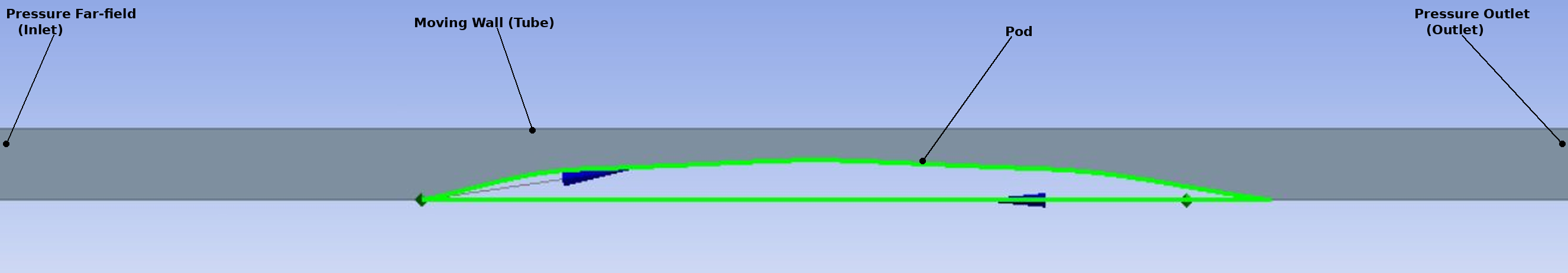}
    \caption{Representative computational domain and boundary-condition implementation for the axisymmetric pod--tube configuration.}
    \label{fig:bc_implementation}
\end{figure*}

\subsubsection{Mesh Generation and Grid Sensitivity}
\label{sec:mesh_generation}

The computational domain is discretised using an unstructured triangular mesh generated in ANSYS Meshing. Inflation layers are applied near the pod surface to resolve boundary-layer gradients. The first-cell height is selected to maintain a near-wall resolution of approximately $y^+ \approx 1$, which is appropriate for resolving the viscous sublayer when using the SST $k$--$\omega$ model. A smooth growth rate is prescribed across the inflation layers to ensure gradual transition from the near-wall mesh to the outer unstructured region.

Mesh quality is monitored using skewness, aspect ratio, and orthogonal quality. Local refinement is applied near the nose, tail, annular throat region, and wake, where strong pressure gradients, flow acceleration, or shock structures may occur.

\begin{figure*}[htbp]
    \centering
    \includegraphics[width=\textwidth]{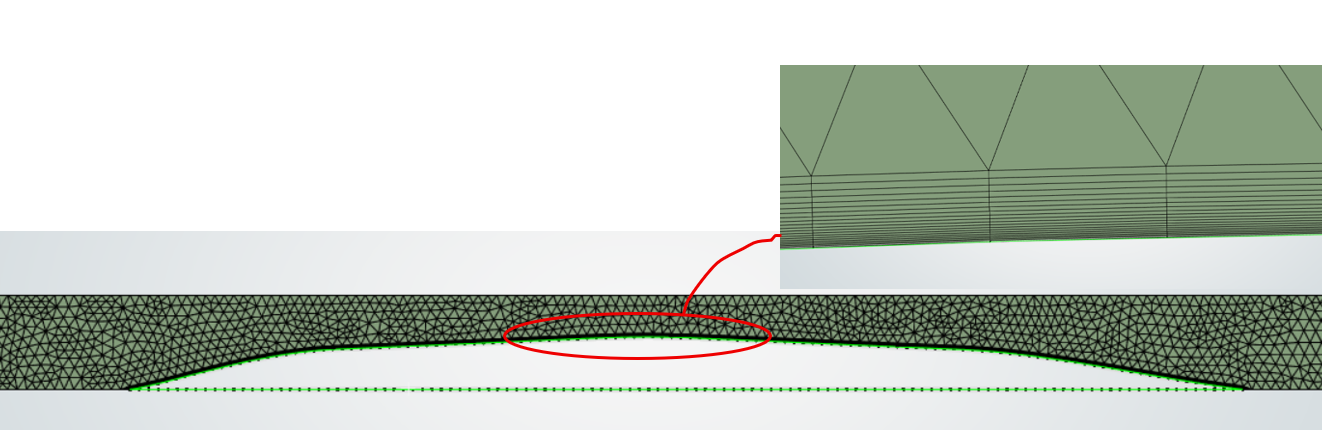}
    \caption{Representative mesh used for the two-dimensional axisymmetric CFD simulations. Inflation layers are applied near the pod surface to resolve the boundary layer.}
    \label{fig:mesh}
\end{figure*}

Grid sensitivity is assessed using representative subsonic and supersonic pod configurations. Three refinement levels, namely coarse, medium, and fine, are generated for each case. The total drag force is used as the primary convergence metric because drag is the target variable for surrogate modelling and optimisation. The results are summarised in Tables~\ref{tab:mesh_subsonic} and~\ref{tab:mesh_supersonic}.

\begin{table}[htbp]
\centering
\small
\renewcommand{\arraystretch}{1.25}
\setlength{\tabcolsep}{6pt}
\begin{tabular}{lcc}
\hline
\textbf{Mesh} & \textbf{Elements} & \textbf{Total drag [N]} \\
\hline
Coarse  & 11{,}266 & 30.164 \\
Medium  & 22{,}440 & 31.690 \\
Fine    & 61{,}780 & 33.718 \\
\hline
\end{tabular}
\caption{Grid sensitivity study for the representative subsonic case.}
\label{tab:mesh_subsonic}
\end{table}

\begin{table}[htbp]
\centering
\small
\renewcommand{\arraystretch}{1.25}
\setlength{\tabcolsep}{6pt}
\begin{tabular}{lcc}
\hline
\textbf{Mesh} & \textbf{Elements} & \textbf{Total drag [N]} \\
\hline
Coarse  & 25{,}935  & 154.94 \\
Medium  & 49{,}454  & 157.42 \\
Fine    & 106{,}157 & 160.42 \\
\hline
\end{tabular}
\caption{Grid sensitivity study for the representative supersonic case.}
\label{tab:mesh_supersonic}
\end{table}

For the supersonic case, the drag variation between the medium and fine meshes is relatively small, indicating acceptable grid convergence for the purposes of large-scale design-space exploration. The subsonic case exhibits a larger medium-to-fine variation, and therefore the mesh-sensitivity results should be interpreted as a compromise between accuracy and computational cost. Since the objective of the present dataset is comparative aerodynamic ranking across many candidate geometries, the medium mesh is adopted for all production simulations. Final optimised geometries are re-evaluated using the CFD solver to verify the surrogate predictions. For future refinement, a formal Grid Convergence Index (GCI) analysis may be used to quantify discretisation uncertainty more rigorously.

\subsubsection{Solver Settings and Turbulence Modelling}
\label{sec:solver_settings}

All simulations are performed using the density-based solver in ANSYS Fluent, which is suitable for compressible flows with strong coupling between pressure, density, and velocity. Second-order upwind discretisation is applied to the flow and turbulence transport equations to reduce numerical diffusion while maintaining numerical stability.

Turbulence closure is provided by the Shear-Stress Transport (SST) $k$--$\omega$ model. This model combines the near-wall accuracy of the $k$--$\omega$ formulation with the freestream robustness of the $k$--$\varepsilon$ model. It is therefore suitable for wall-bounded compressible flows involving adverse pressure gradients, boundary-layer development, and possible separation near the tail or in the annular passage. The use of near-wall resolution with $y^+ \approx 1$ allows the viscous sublayer to be resolved directly rather than modelled using wall functions.

\subsubsection{CFD Validation}
\label{sec:cfd_validation}

The CFD methodology is validated against the numerical study of Jang et al.~\cite{Jang2021}, which reports axisymmetric compressible-flow simulations of a Hyperloop pod inside an evacuated tube using ANSYS Fluent and the SST $k$--$\omega$ turbulence model. The operating conditions from the reference study are reproduced in the present setup. Although Jang et al.\ used a moving overset-mesh formulation, the present pod-fixed reference frame imposes the same relative motion between pod, tube wall, and surrounding gas.

\begin{figure}[htbp]
    \centering
    \includegraphics[width=\linewidth]{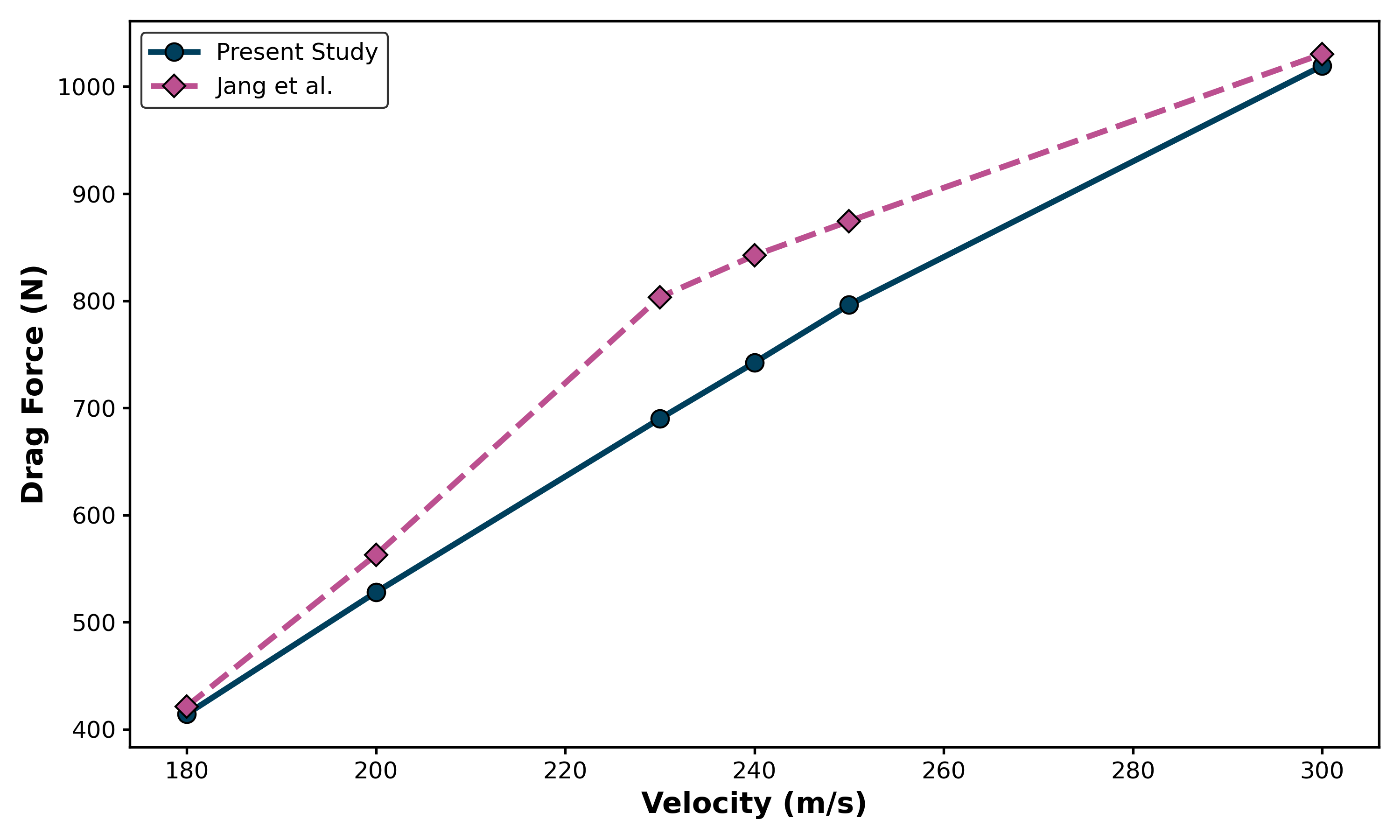}
    \caption{Validation of the present CFD setup against the numerical results of Jang et al.~\cite{Jang2021}.}
    \label{fig:drag_validation}
\end{figure}

A velocity sweep is simulated, and the resulting total drag values are compared with the published data, as shown in Figure~\ref{fig:drag_validation}. The present CFD setup shows close agreement at the primary operating velocity, with a discrepancy of approximately 1\% at 300\,m/s. Larger deviations in the 230--240\,m/s range are attributed to the sensitivity of the flow near choking onset, where small differences in mesh topology, boundary treatment, and reference-frame implementation can produce amplified differences in integrated drag.

The validation exercise confirms that the present CFD setup captures the dominant compressible-flow behaviour in an evacuated-tube environment with sufficient accuracy for dataset generation and comparative aerodynamic optimisation.

\subsection{Dataset Generation and Preparation}
\label{sec:dataset_preparation}

The CFD simulations generate the dataset used for surrogate-model training. Each geometry is represented by a vector of design variables,

\begin{equation}
    \mathbf{x} =
    \left[
    l_\text{nose},
    l_\text{body},
    l_\text{tail},
    s_\text{pos},
    \alpha_\text{nose},
    \alpha_\text{tail},
    h_j
    \right],
    \label{eq:design_vector}
\end{equation}

\noindent where $l_\text{nose}$, $l_\text{body}$, and $l_\text{tail}$ denote the nose, body, and tail lengths; $s_\text{pos}$ denotes the shoulder position; $\alpha_\text{nose}$ and $\alpha_\text{tail}$ denote the nose and tail angles; and $h_j$ denotes the junction radius.

For each sampled geometry, the validated CFD solver is used to compute the aerodynamic force components on the pod surface. The recorded quantities include total drag, pressure drag, and viscous drag. Total drag is used as the primary surrogate target and optimisation objective, while the pressure and viscous components are retained for physical interpretation of the optimised designs.

The dataset consists of input--output pairs of the form

\begin{equation}
    \mathcal{D} =
    \left\{
    \left(\mathbf{x}^{(i)}, D_\text{total}^{(i)}\right)
    \right\}_{i=1}^{N},
    \label{eq:dataset}
\end{equation}

\noindent where $\mathbf{x}^{(i)}$ is the design-vector representation of the $i$th pod geometry, $D_\text{total}^{(i)}$ is the corresponding CFD-computed total drag, and $N$ is the number of evaluated configurations.

The dataset is divided into training and test subsets using an 80:20 split. The training set is used to fit the surrogate model, while the held-out test set is used only for evaluating generalisation performance. The optimisation search is restricted to the same parameter bounds used for dataset generation, ensuring that the surrogate is used primarily in interpolation rather than extrapolation.

\subsection{Surrogate Model Development}
\label{sec:surrogate_model}

A machine-learning surrogate model is trained to approximate the mapping between pod geometry and total aerodynamic drag. Once trained, the surrogate replaces the CFD solver inside the optimisation loop, reducing the cost of each design evaluation from a full CFD simulation to a rapid model prediction.

Prior to model training, second-degree polynomial features are generated from the original geometric inputs to represent nonlinear interactions among design variables. The resulting feature set is standardised using zero-mean, unit-variance scaling. The transformation is fitted using the training data and then applied to the test data to avoid information leakage.

Several regression models are benchmarked using the same train--test split: Linear Regression, Random Forest, Gradient Boosting, and XGBoost. Model performance is evaluated using the coefficient of determination $R^2$ and error metrics including mean absolute error (MAE), mean squared error (MSE), and root-mean-square error (RMSE). In addition, the generalisation gap, defined as the difference between training $R^2$ and test $R^2$, is used to assess overfitting.

The benchmark results are summarised in Table~\ref{tab:model_benchmark}. Linear Regression shows insufficient predictive capacity for the nonlinear geometry--drag relationship. Random Forest achieves high training accuracy but exhibits a larger generalisation gap. Gradient Boosting improves test accuracy, while XGBoost provides the best overall balance of test performance and generalisation. XGBoost is therefore selected as the final surrogate model.

\begin{table}[htbp]
\centering
\small
\renewcommand{\arraystretch}{1.25}
\setlength{\tabcolsep}{6pt}
\begin{tabular}{lccc}
\hline
\textbf{Algorithm} & \textbf{Train $R^2$} & \textbf{Test $R^2$} & \textbf{Gap} \\
\hline
Linear Regression & 0.6601 & 0.6430 & 0.0171 \\
Random Forest     & 0.9821 & 0.8164 & 0.1657 \\
Gradient Boosting & 0.9876 & 0.9436 & 0.0440 \\
XGBoost           & 0.9743 & 0.9524 & 0.0219 \\
\hline
\end{tabular}
\caption{Regression-model benchmark based on training accuracy, test accuracy, and generalisation gap.}
\label{tab:model_benchmark}
\end{table}

XGBoost builds an additive ensemble of regression trees. At boosting iteration $t$, the prediction for sample $i$ is updated as

\begin{equation}
    \hat{y}_i^{(t)} =
    \hat{y}_i^{(t-1)} + \eta f_t(\mathbf{x}_i),
    \label{eq:xgb_update}
\end{equation}

\noindent where $\eta$ is the learning rate and $f_t$ is the newly added regression tree. The tree is selected by minimising a regularised objective function,

\begin{equation}
    \mathcal{L}^{(t)} =
    \sum_{i=1}^{m}
    l\left(
    y_i,\,
    \hat{y}_i^{(t-1)} + f_t(\mathbf{x}_i)
    \right)
    + \Omega(f_t),
    \label{eq:xgb_objective}
\end{equation}

\noindent where $l$ is the loss function and $\Omega(f_t)$ penalises model complexity. The regularisation term is expressed as

\begin{equation}
    \Omega(f_t) =
    \gamma T +
    \frac{1}{2}\lambda \sum_{j=1}^{T} w_j^2,
    \label{eq:xgb_regularisation}
\end{equation}

\noindent where $T$ is the number of leaves in the tree, $w_j$ is the score assigned to leaf $j$, and $\gamma$ and $\lambda$ are regularisation parameters.

The final XGBoost configuration uses shallow trees, a low learning rate, row and column subsampling, and regularisation to control overfitting. These choices promote gradual ensemble growth, reduce variance, and improve robustness on the relatively small CFD-generated dataset.

\subsection{Optimisation Framework}
\label{sec:optimisation_framework}

The trained XGBoost surrogate is coupled with two population-based global optimisation algorithms: Differential Evolution (DE) and Particle Swarm Optimisation (PSO). Both algorithms search the bounded geometric design space for the pod configuration that minimises predicted total drag.

The optimisation problem is formulated as

\begin{equation}
    \min_{\mathbf{x}} \quad
    f(\mathbf{x}) =
    \hat{D}_\text{total}(\mathbf{x}),
    \label{eq:opt_objective}
\end{equation}

\noindent subject to

\begin{equation}
    x_j^\text{lb}
    \leq
    x_j
    \leq
    x_j^\text{ub},
    \qquad
    j = 1,2,\ldots,7,
    \label{eq:opt_bounds}
\end{equation}

\noindent where $\hat{D}_\text{total}$ is the drag predicted by the surrogate model, and $x_j^\text{lb}$ and $x_j^\text{ub}$ are the lower and upper bounds of the $j$th design variable. The bounds are identical to those used during LHS-based dataset generation, thereby avoiding surrogate extrapolation outside the trained design space.

\subsubsection{Differential Evolution}
\label{sec:de}

Differential Evolution evolves a population of candidate design vectors using mutation, crossover, and selection. For each target vector $\mathbf{x}_i$, a mutant vector is generated as

\begin{equation}
    \mathbf{v}_i =
    \mathbf{x}_{r_1}
    +
    F\left(
    \mathbf{x}_{r_2}
    -
    \mathbf{x}_{r_3}
    \right),
    \label{eq:de_mutation}
\end{equation}

\noindent where $r_1$, $r_2$, and $r_3$ are distinct randomly selected population indices, and $F$ is the mutation scale factor. A trial vector is then formed through binomial crossover,

\begin{equation}
    u_{i,j} =
    \begin{cases}
        v_{i,j}, & \mathrm{if}\;\; r_j \leq CR, \\
        x_{i,j}, & \mathrm{otherwise},
    \end{cases}
    \label{eq:de_crossover}
\end{equation}

\noindent where $CR$ is the crossover probability. The trial vector replaces the target vector only if it produces a lower surrogate-predicted drag. In this study, DE is run with a population size of 30 over 100 iterations. The \texttt{polish} option is enabled to perform a local refinement step after global convergence.

\subsubsection{Particle Swarm Optimisation}
\label{sec:pso}

Particle Swarm Optimisation represents each candidate solution as a particle moving through the design space. Each particle updates its position based on its current velocity, its own best-known position, and the best-known position of the swarm. The velocity and position updates are given by

\begin{equation}
    \mathbf{v}_i^{t+1}
    =
    w\mathbf{v}_i^t
    +
    c_1 r_1
    \left(
    \mathbf{p}_i - \mathbf{x}_i^t
    \right)
    +
    c_2 r_2
    \left(
    \mathbf{g} - \mathbf{x}_i^t
    \right),
    \label{eq:pso_velocity}
\end{equation}

\begin{equation}
    \mathbf{x}_i^{t+1}
    =
    \mathbf{x}_i^t
    +
    \mathbf{v}_i^{t+1},
    \label{eq:pso_position}
\end{equation}

\noindent where $w$ is the inertia weight, $c_1$ and $c_2$ are the cognitive and social acceleration coefficients, $r_1$ and $r_2$ are random numbers, $\mathbf{p}_i$ is the personal-best position of particle $i$, and $\mathbf{g}$ is the global-best position of the swarm.

In this study, PSO is configured with 30 particles evolved over 100 iterations. The inertia weight is set to $w=0.5$, and the acceleration coefficients are set to $c_1=c_2=1.5$, providing a balance between design-space exploration and convergence toward the best solution identified by the swarm.

\subsubsection{Selection and CFD Verification of the Optimal Design}
\label{sec:optimal_design_selection}

DE and PSO are executed independently using the same surrogate objective function and identical geometric bounds. The candidate design producing the lowest surrogate-predicted total drag is selected as the optimal geometry for the corresponding operating regime. This design is reconstructed using the B\'{e}zier parameterisation and then re-evaluated using the CFD solver.

This final CFD re-evaluation serves two purposes. First, it verifies that the surrogate-predicted drag reduction is consistent with the full-order CFD solution. Second, it confirms that the optimised geometry does not exploit surrogate artefacts or interpolation errors in the design space. The agreement between surrogate prediction and CFD recomputation is therefore used as the final validation step for the optimisation workflow.
\section{Converging--Diverging Tube Concept for Supersonic Hyperloop Transport}
\label{sec:cd_tube_concept}

A central aerodynamic limitation in supersonic evacuated tube transport is the onset of flow choking in the annular passage between the pod surface and the tube wall. In the pod-fixed reference frame, the incoming air is forced through a restricted area around the pod. As the pod Mach number and blockage ratio increase, the annular flow can locally accelerate to sonic conditions. Once this occurs, the mass flow rate through the minimum-area region becomes constrained, causing upstream pressure buildup, shock formation, and a rapid increase in aerodynamic drag. This choking constraint limits the attainable operating envelope and must be addressed for any practical supersonic ETT architecture.

The theoretical framework in Section~\ref{sec:theoretical_framework} shows that the feasibility of supersonic operation depends not only on the final operating Mach number and blockage ratio, but also on the path by which this state is reached. In particular, the region between the isentropic and Kantrowitz limits can admit multiple flow states, where the realized solution depends on whether the system enters from an unchoked or choked operating branch. This motivates the use of staged pressure--area management as a potential strategy for maintaining an unchoked flow path during acceleration to supersonic cruise.

\subsection{Proposed Staged Pressure--Area Management Concept}
\label{sec:staged_cd_concept}

This study proposes a staged converging--diverging (CD) tube concept in which the tube cross-sectional area and internal static pressure are varied in a coordinated manner along the direction of travel. The objective is not to operate the tube as a conventional de Laval nozzle, but to use controlled area variation and pressure staging to manage the effective blockage ratio and compressible-flow response around the moving pod.

\begin{figure*}[htbp]
    \centering
    \includegraphics[width=\textwidth]{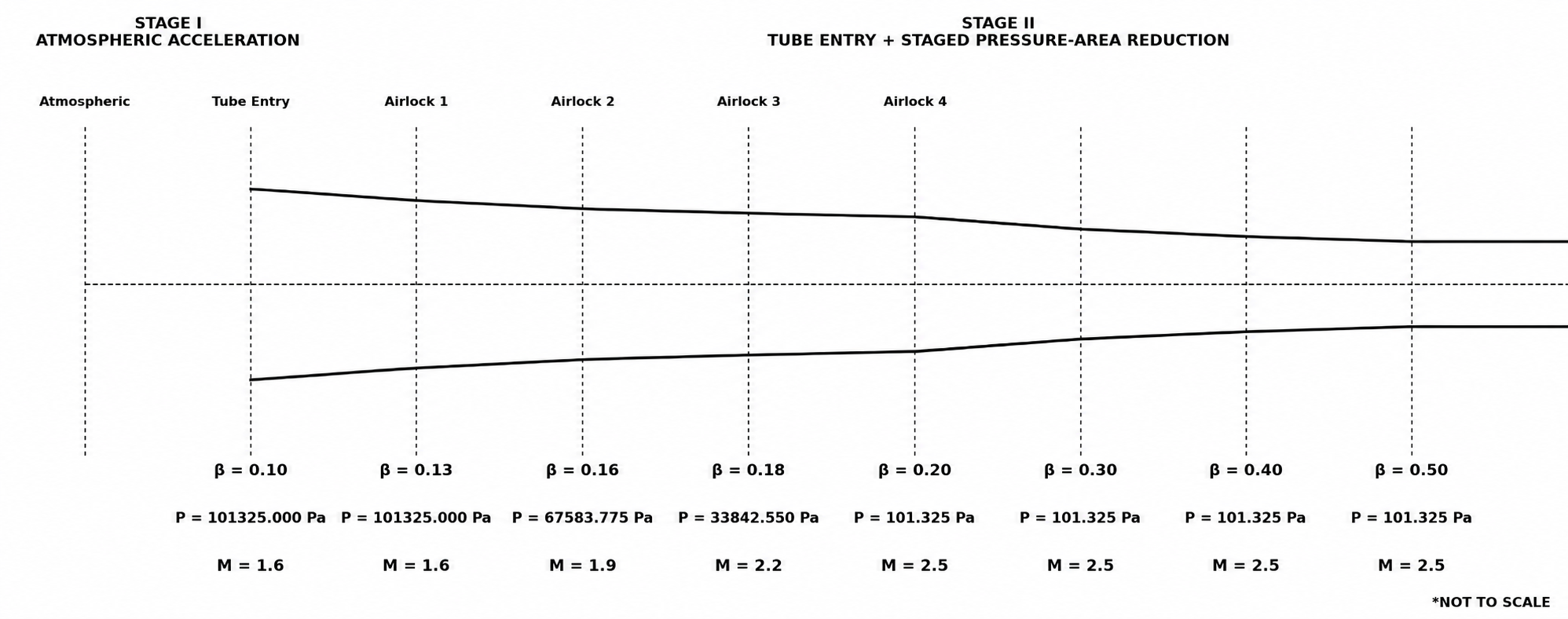}
    \caption{Schematic of the proposed staged converging--diverging tube concept for supersonic ETT operation. The tube-area variation and pressure staging are coordinated to control the effective blockage ratio and reduce the likelihood of annular-flow choking during acceleration to near-vacuum cruise.}
    \label{fig:cd_concept}
\end{figure*}

The concept consists of three operational phases: atmospheric acceleration, staged tube entry with pressure reduction, and staged deceleration with pressure recovery. A schematic representation of the proposed layout is shown in Figure~\ref{fig:cd_concept}. The figure indicates the qualitative variation of tube area, blockage ratio $\beta$, static pressure, and pod Mach number along the acceleration and cruise portions of the trajectory.

\subsubsection{Stage I: Atmospheric Acceleration Before Tube Entry}

In the first stage, the pod is accelerated outside the evacuated tube to an initial supersonic speed of Mach~1.6 under atmospheric pressure,
$P = 101{,}325~\mathrm{Pa}$. Since the pod is not yet confined by the tube, the effective blockage ratio is $\beta = 0$. This stage avoids the need to accelerate the pod from rest to supersonic speed inside a high-pressure confined tube, where wave drag, viscous losses, and choking constraints would be severe.

However, atmospheric supersonic acceleration introduces its own practical limitations. At Mach~1.6, the pod generates an oblique shock system that can propagate to the ground and produce a sonic-boom signature. Therefore, this stage should be interpreted as a conceptual acceleration strategy rather than a complete operational solution. In a practical implementation, additional sonic-boom mitigation measures would be required.

One possible passive mitigation approach is the use of inclined side walls or shock-deflection panels along the acceleration track. These surfaces may intercept outward-propagating oblique shocks and redirect part of the wave energy away from the ground-level surroundings. The concept is analogous in broad purpose to noise barriers used in high-speed rail corridors, but the design objective is different: instead of attenuating broadband acoustic noise, the walls would be shaped to manage shock propagation. A detailed assessment of such shock-deflection infrastructure is outside the scope of the present study and would require dedicated unsteady three-dimensional simulations and acoustic analysis.

\subsubsection{Stage II: Tube Entry with Coordinated Pressure and Area Reduction}

After atmospheric acceleration, the pod enters a large-diameter tube section with a low initial blockage ratio. In the present concept, the entry section begins at approximately $\beta = 0.10$, placing the flow within an unchoked operating regime. From this point, the tube cross-section is reduced gradually along the direction of motion, thereby increasing the effective blockage ratio toward the intended cruise value.

At the same time, the static pressure inside the tube is reduced in stages using airlock chambers and vacuum pumping. This pressure reduction lowers the gas density and therefore reduces the absolute aerodynamic load on the pod. More importantly, the staged transition avoids imposing an abrupt pressure discontinuity between atmospheric entry and near-vacuum cruise conditions, which could otherwise generate strong transient shocks and large unsteady pressure loads.

The proposed entry strategy therefore combines two coupled mechanisms:

\begin{enumerate}
    \item \textbf{Geometric convergence:} The tube cross-section decreases progressively, increasing the blockage ratio from the initial low-blockage entry value toward the final cruise value. The operating path is selected so that the annular flow remains below the choking boundary predicted by the isentropic limit.

    \item \textbf{Staged pressure reduction:} The static pressure is reduced through a sequence of controlled airlock stages, allowing the aerodynamic environment to transition gradually from atmospheric entry to near-vacuum cruise. This reduces the risk of strong shock formation associated with sudden back-pressure changes.
\end{enumerate}

Within the idealized framework considered here, the combined pressure--area staging is intended to keep the flow on the unchoked branch while the pod transitions toward higher Mach number and higher blockage ratio. This interpretation is consistent with the path-dependent operating behaviour discussed in Section~\ref{sec:theoretical_framework}. The strategy should therefore be viewed as a proposed aerodynamic operating pathway rather than a complete engineering design of the tube, airlock, and pumping system.

\subsubsection{Stage III: Diverging Tube Section and Pressure Recovery During Deceleration}

The deceleration phase follows the reverse logic of the entry process. As the pod approaches its destination, the tube transitions into a diverging section, increasing the local tube cross-sectional area and reducing the effective blockage ratio. This expansion reduces aerodynamic confinement and helps avoid the formation of choked regions during deceleration.

Simultaneously, the static pressure is restored in stages through controlled air admission in successive airlock chambers. Gradual pressure recovery avoids an abrupt rise in back pressure, which could generate strong reflected shocks and large unsteady pressure gradients near the pod. The pod then exits the tube at near-atmospheric conditions and completes the final deceleration phase externally or within a dedicated terminal section.

The present study focuses primarily on the acceleration and cruise portions of this concept. Detailed design of the deceleration section, airlock timing, vacuum-pump capacity, structural response of the tube, and passenger-comfort constraints associated with acceleration and deceleration are beyond the scope of the current aerodynamic analysis.

\subsection{Numerical Assessment of the CD Tube Concept}
\label{sec:cd_numerical_assessment}

The proposed CD tube concept is evaluated using a sequence of steady axisymmetric CFD simulations corresponding to representative operating points along the staged acceleration process. The purpose of these simulations is to assess whether the prescribed pressure--area path can maintain an unchoked annular flow in the idealized computational framework adopted in this study.

The numerical assessment begins with the pod operating at Mach~1.6 under atmospheric conditions, which provides the pre-entry aerodynamic baseline. The tube-entry phase is then represented by a sequence of operating points beginning at a low blockage ratio of $\beta = 0.10$. Across subsequent stages, the blockage ratio is increased through geometric convergence of the tube, while the static pressure is reduced toward the near-vacuum cruise condition. Each operating point is checked against the theoretical choking limits discussed in Section~\ref{sec:theoretical_framework}, and the CFD solution is examined for evidence of sonic blockage, upstream pressure buildup, and shock-induced flow separation in the annular passage.

The final simulated operating condition corresponds to Mach~2.5 and blockage ratio $\beta = 0.50$ under near-vacuum pressure. Within the assumptions of the steady two-dimensional axisymmetric model, the CFD results indicate that the annular flow remains unchoked along the selected operating path. The drag also decreases substantially as the pressure is reduced, reflecting the transition from wave-drag-dominated atmospheric supersonic operation to a much lower-density near-vacuum cruise regime.

To reduce the possibility that these trends are numerical artefacts, selected operating points are recomputed using refined meshes and identical convergence criteria. Residual histories and integrated force monitors are also tracked to ensure that each operating point reaches a steady solution before the next stage is analysed. These checks support the numerical consistency of the observed pressure and drag trends, although they do not replace a full transient simulation of the complete acceleration process.

The results of this numerical assessment are presented in Section~\ref{SUPERSONICVALIDATION}. They should be interpreted as a preliminary aerodynamic feasibility study of staged pressure--area management, rather than as a complete validation of a deployable supersonic Hyperloop system.

\section{Results and Discussion}
\label{sec:results}

This section presents the aerodynamic performance of the optimized pod geometries obtained from the CFD--ML optimization framework. The discussion is divided into two parts. First, the optimized subsonic pod is analyzed in terms of flow structure, pressure distribution, wall shear stress, drag decomposition, and surrogate-model accuracy. Second, the supersonic case is examined through the proposed staged CD tube operating sequence, including atmospheric baseline operation, tube entry, staged pressure reduction, near-vacuum cruise, and blockage-ratio effects.

\subsection{Subsonic Case Study}
\label{sec:subsonic_results}

\begin{figure*}[t]
    \centering
    \includegraphics[width=\textwidth]{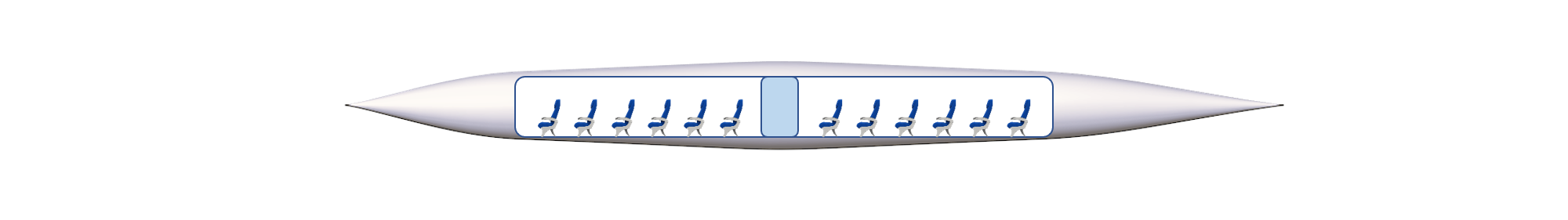}
    \caption{Optimized pod geometry obtained for the subsonic operating condition.}
    \label{fig:subsonicoptimized}
\end{figure*}

The optimized subsonic geometry identified by the Differential Evolution--XGBoost optimization framework is shown in Figure~\ref{fig:subsonicoptimized}. This case is evaluated at a freestream Mach number of $M_\infty = 0.42$ and a tube static pressure of $101.325$\,Pa, corresponding to the low-pressure operating environment considered for subsonic Hyperloop operation. The objective of this analysis is to verify the aerodynamic behavior of the optimized geometry using CFD and to assess the accuracy of the surrogate prediction at the final design point.

\subsubsection{Mach Number and Pressure Fields}
\label{sec:subsonic_flowfield}

\begin{figure*}[!t]
    \centering
    \begin{subfigure}{0.85\textwidth}
        \centering
        \includegraphics[width=\textwidth]{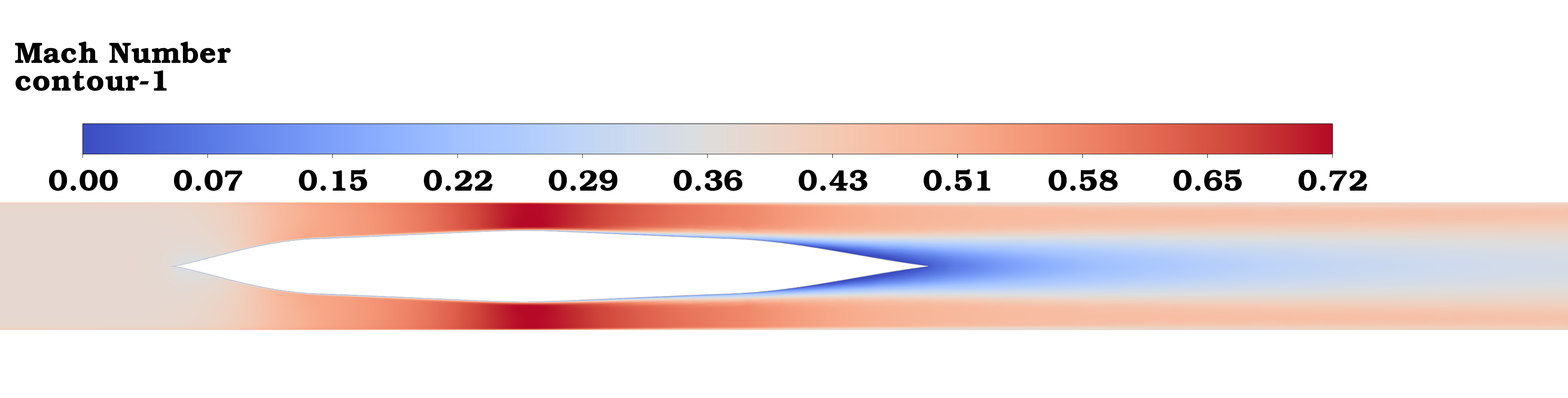}
        \caption{Mach number contour for the optimized subsonic pod.}
        \label{fig:sub_mach}
    \end{subfigure}

    \vspace{0.35cm}

    \begin{subfigure}{0.85\textwidth}
        \centering
        \includegraphics[width=\textwidth]{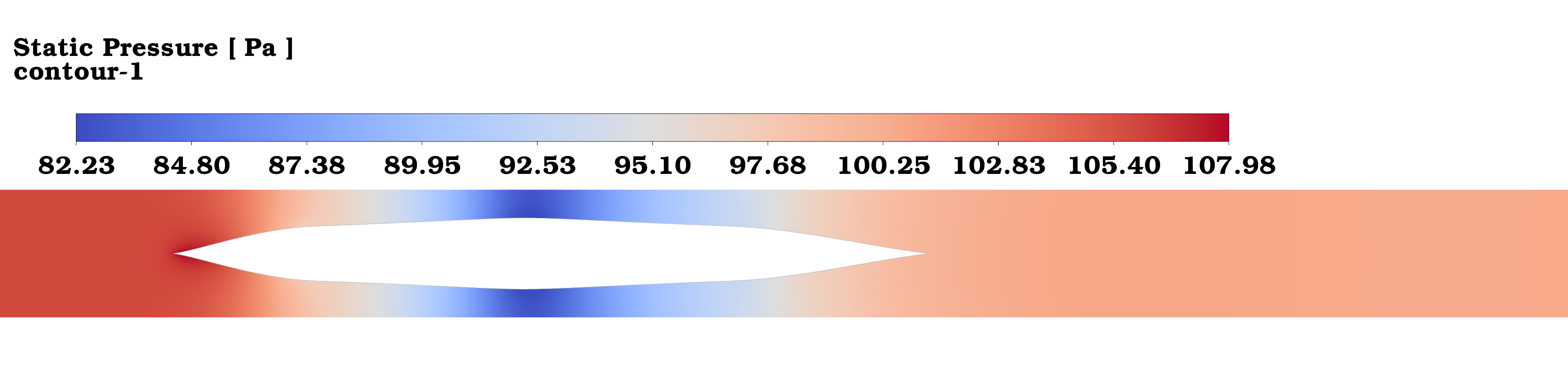}
        \caption{Static pressure contour for the optimized subsonic pod.}
        \label{fig:sub_pressure}
    \end{subfigure}

    \caption{Flow-field structure around the optimized subsonic pod.}
    \label{fig:subsonic_contours}
\end{figure*}

The Mach number contour in Figure~\ref{fig:sub_mach} shows the expected acceleration and deceleration pattern of confined subsonic flow around a slender axisymmetric body. The incoming flow at $M_\infty = 0.42$ decelerates near the nose stagnation region and then accelerates through the annular passage as the pod radius increases. The maximum local Mach number reaches approximately $M \approx 0.72$ near the maximum-radius section, indicating a local acceleration of nearly 75\% relative to the freestream Mach number. This increase is caused by the reduction in effective flow area between the pod surface and the tube wall.

Downstream of the maximum-radius region, the pod profile tapers toward the tail and the annular passage area increases. As a result, the flow decelerates and partially recovers toward the freestream condition. No shock structure or local sonic region is observed, confirming that the optimized pod remains fully subsonic under the specified operating condition.

The static pressure field in Figure~\ref{fig:sub_pressure} is consistent with the Mach number distribution. A localized high-pressure region forms near the nose, where the flow decelerates, with a peak static pressure of approximately $107.98$\,Pa. The pressure then decreases along the forebody as the flow accelerates through the annular gap, reaching a minimum of approximately $82.23$\,Pa near the maximum-radius section. This pressure difference between the forebody and aft region produces the main pressure-drag contribution.

Along the tail, the static pressure partially recovers as the body radius decreases and the flow decelerates. The smooth recovery indicates that the optimized tail profile avoids abrupt adverse pressure gradients. The far-field pressure remains close to the prescribed tube pressure, suggesting that the upstream and downstream domain extents are sufficient to prevent significant boundary-induced distortion of the local pod flow.

\subsubsection{Wall Shear Stress and Separation Behavior}
\label{sec:subsonic_wss}

\begin{figure*}[t]
    \centering
    \includegraphics[width=0.7\textwidth]{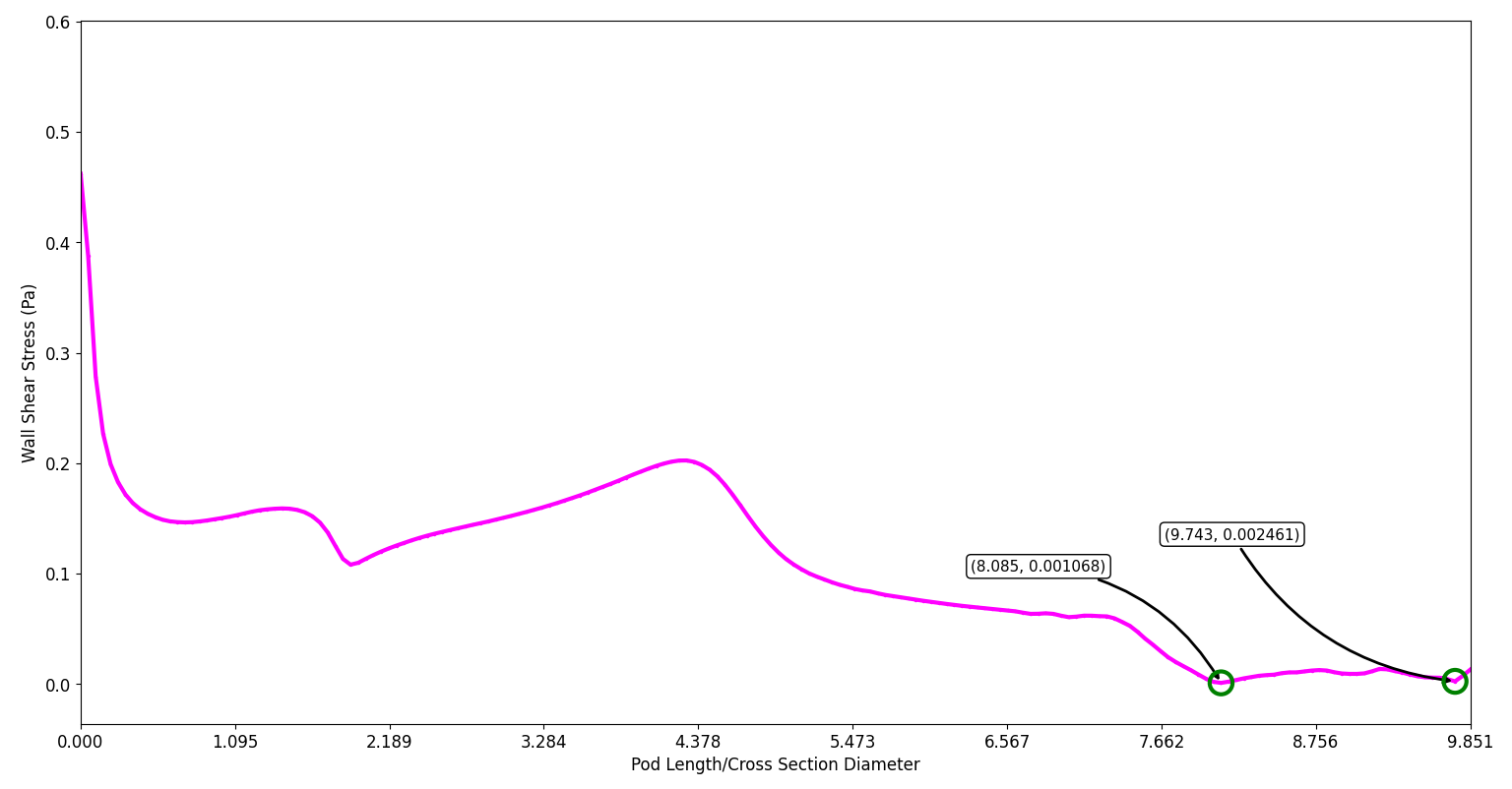}
    \caption{Wall shear stress distribution along the normalized pod length for the optimized subsonic geometry.}
    \label{fig:wss}
\end{figure*}

The wall shear stress distribution along the pod surface is shown in Figure~\ref{fig:wss}. The shear stress is largest near the forebody region, where the boundary layer develops under strong local acceleration. A peak value of approximately $0.45$\,Pa occurs close to the nose region, followed by a rapid reduction along the forebody. A secondary increase to approximately $0.20$\,Pa is observed near the maximum-radius section, consistent with the local peak in Mach number and the associated increase in near-wall velocity gradient.

Beyond the shoulder, the wall shear stress decreases gradually toward the tail. Importantly, the wall shear stress remains positive over the full pod length. Since negative wall shear stress would indicate local flow reversal, the absence of such a region suggests that the optimized tail suppresses boundary-layer separation under the subsonic operating condition. This is consistent with the smooth pressure recovery observed in Figure~\ref{fig:sub_pressure}.

\subsubsection{Drag Decomposition and Surrogate Verification}
\label{sec:subsonic_drag}

The CFD-computed total drag for the optimized subsonic pod is

\begin{equation}
    D_{\text{CFD}} = 31.69~\mathrm{N}.
\end{equation}

Pressure drag accounts for approximately 59.9\% of the total drag, while viscous drag contributes the remaining 40.1\%. The larger pressure-drag fraction is consistent with the confined-tube environment, where the blockage-induced pressure difference between the nose and tail remains an important resistance mechanism even at low pressure. The viscous contribution is also significant because of the large wetted area and the turbulent boundary layer resolved using the SST $k$--$\omega$ model.

For the same optimized geometry, the XGBoost surrogate predicts

\begin{equation}
    D_{\text{surrogate}} = 32.19~\mathrm{N}.
\end{equation}

The relative discrepancy between the surrogate prediction and CFD recomputation is therefore

\begin{equation}
    \epsilon =
    \frac{
    \left|D_{\text{surrogate}} - D_{\text{CFD}}\right|
    }{
    D_{\text{CFD}}
    }
    \times 100
    =
    1.58\%.
\end{equation}

This agreement indicates that the surrogate model provides an accurate prediction at the optimized design point. More importantly, the optimized geometry does not appear to exploit a surrogate artifact, since the independent CFD recomputation confirms the low-drag performance predicted during the optimization stage.

\subsection{Supersonic Case Study}
\label{SUPERSONICVALIDATION}

The supersonic analysis focuses on two related objectives. The first is to compare optimized pod geometries obtained under open-atmosphere and confined near-vacuum operating conditions. The second is to assess the proposed staged CD tube concept as a pressure--area management strategy for reaching near-vacuum supersonic cruise while avoiding the formation of a choked annular-flow region in the idealized axisymmetric simulations.

\subsubsection{Selection of the Supersonic Reference Geometry}
\label{sec:supersonic_geometry_selection}

\begin{figure*}[!t]
    \centering
    \begin{subfigure}{\textwidth}
        \centering
        \includegraphics[width=\textwidth]{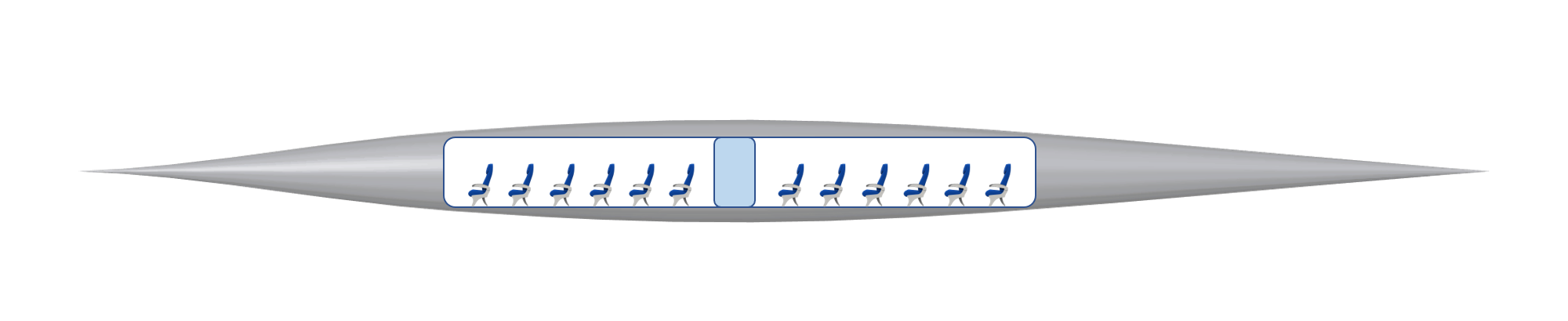}
        \caption{Optimized geometry for confined near-vacuum operation at $M_\infty = 2.5$, $P_\infty = 101.325$\,Pa, and $\beta = 0.20$.}
        \label{fig:pod_front}
    \end{subfigure}

    \vspace{0.3cm}

    \begin{subfigure}{\textwidth}
        \centering
        \includegraphics[width=\textwidth]{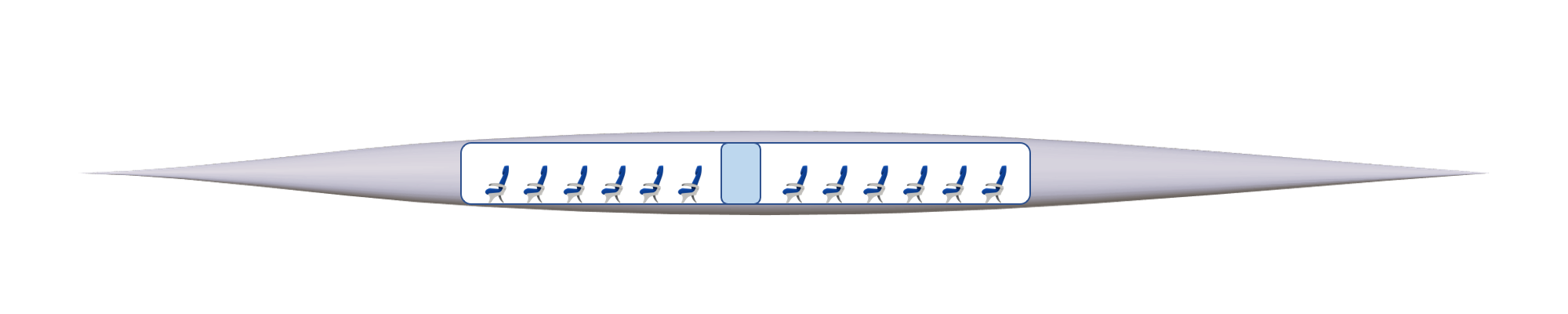}
        \caption{Optimized geometry for open-atmosphere operation at $M_\infty = 1.6$ and $P_\infty = 101{,}325$\,Pa.}
        \label{fig:pod_side}
    \end{subfigure}

    \caption{Supersonic pod geometries optimized for different operating environments.}
    \label{fig:pod_geometry}
\end{figure*}

Two independent supersonic optimization studies are performed. The first corresponds to open-atmosphere operation at $M_\infty = 1.6$ and $P_\infty = 101{,}325$\,Pa. The second corresponds to confined-tube operation at $M_\infty = 2.5$, $P_\infty = 101.325$\,Pa, and $\beta = 0.20$. The resulting geometries are shown in Figure~\ref{fig:pod_geometry}.

When the atmosphere-optimized geometry is evaluated under confined-tube conditions, it produces higher drag than the geometry optimized directly for the near-vacuum tube environment. This confirms that the two operating regimes require different aerodynamic compromises. In open atmosphere, the design is primarily governed by wave drag and forebody shock structure. In contrast, the confined-tube case is affected by annular-flow acceleration, wall proximity, pressure recovery, and viscous shear in the narrow passage between the pod and tube wall. The tube-optimized geometry is therefore used as the reference geometry for the staged CD tube simulations.

The drag coefficient is defined as

\begin{equation}
    C_D =
    \frac{D}
    {\tfrac{1}{2}\gamma P_\infty M_\infty^2 A_{\text{ref}}},
    \qquad
    A_{\text{ref}} = \pi(1.2)^2 = 4.524~\mathrm{m}^2,
    \label{eq:cd_definition}
\end{equation}

\noindent where $D$ is the total drag, $P_\infty$ is the static pressure, $M_\infty$ is the freestream Mach number, $\gamma = 1.4$, and $A_{\text{ref}}$ is the pod frontal area. The operating conditions and corresponding drag values for the staged CD tube simulations are summarized in Table~\ref{tab:cd_tube_stages}.

\begin{table*}[!t]
\centering
\small
\renewcommand{\arraystretch}{1.3}
\setlength{\tabcolsep}{6pt}
\begin{tabular}{lcccc}
\hline
\textbf{Stage} & \textbf{$\beta$} & \textbf{$M_\infty$}
               & \textbf{Tube pressure [Pa]}
               & \textbf{Total drag [N]} \\
\hline
Stage I: atmospheric baseline      & ---  & 1.6 & 101{,}325.000 & 68{,}913.7 \\
Stage II: tube entry               & 0.10 & 1.6 & 101{,}325.000 & 22{,}479.1 \\
Phase 1                            & 0.13 & 1.6 & 101{,}325.000 & 19{,}998.1 \\
Phase 2                            & 0.16 & 1.9 & 67{,}583.775  & 21{,}893.7 \\
Phase 3                            & 0.18 & 2.2 & 33{,}842.550  & 16{,}384.1 \\
Phase 4                            & 0.20 & 2.5 & 101.325       & 123.6 \\
Phase 5                            & 0.30 & 2.5 & 101.325       & 133.9 \\
Phase 6                            & 0.40 & 2.5 & 101.325       & 125.9 \\
Phase 7                            & 0.50 & 2.5 & 101.325       & 120.5 \\
\hline
\end{tabular}
\caption{Operating conditions and computed total drag values for the staged CD tube simulation sequence.}
\label{tab:cd_tube_stages}
\end{table*}

\subsubsection{Atmospheric Supersonic Baseline}
\label{sec:atmospheric_baseline}

The atmospheric baseline corresponds to open operation at $M_\infty = 1.6$ and $P_\infty = 101{,}325$\,Pa. Since the pod is not confined by a tube in this case, the effective blockage ratio is zero and no annular choking constraint is present. The simulation therefore provides a reference for the aerodynamic loads associated with supersonic operation before tube entry.

At this condition, the pod generates an oblique shock system from the forebody, followed by expansion and recompression features along the body and tail. Because the flow is not confined by a tube wall, the shock system propagates outward and the wake remains characterized by low pressure and incomplete downstream recovery. The resulting total drag is

\begin{equation}
    D = 68{,}913.7~\mathrm{N},
\end{equation}

\noindent corresponding to $C_D = 0.0839$ using Eq.~\ref{eq:cd_definition}. This is the largest drag value in the present study and serves as the upper-bound reference for comparison with the confined and low-pressure operating stages.

\subsubsection{Tube Entry and Staged Pressure Reduction}
\label{sec:tube_entry_results}

\begin{figure}[!t]
\centering
\begin{subfigure}{\columnwidth}
\centering
\includegraphics[width=\linewidth]{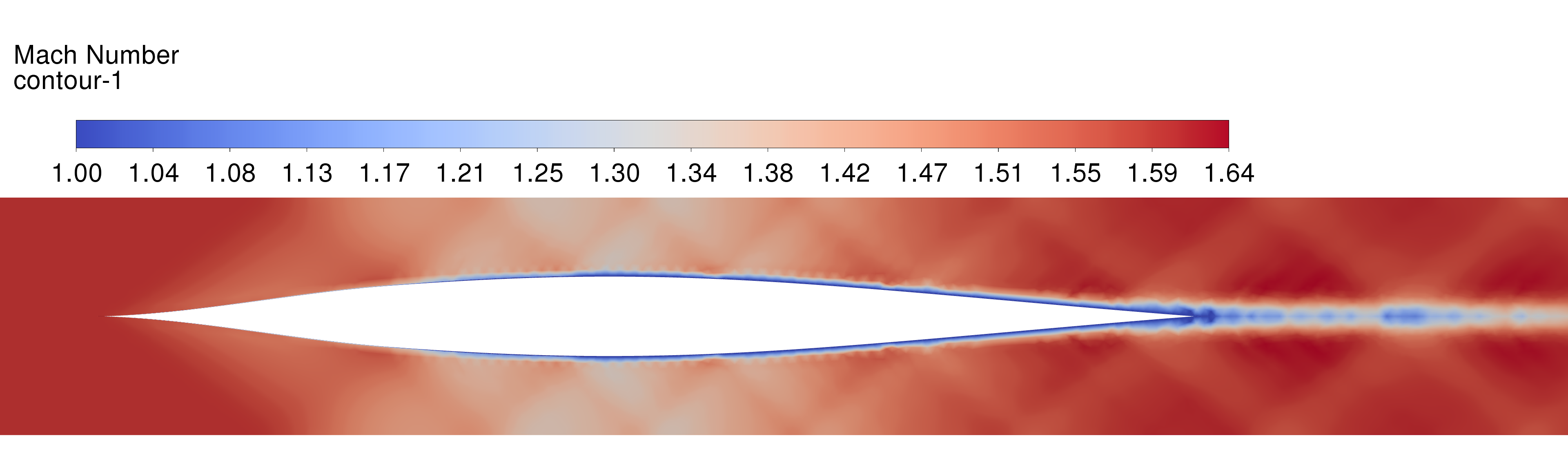}
\caption{Phase 1: $\beta=0.13$, $M_\infty=1.6$, $P_\infty=101325$ Pa.}
\label{fig:ss_s1_mach}
\end{subfigure}

\vspace{0.15cm}

\begin{subfigure}{\columnwidth}
\centering
\includegraphics[width=\linewidth]{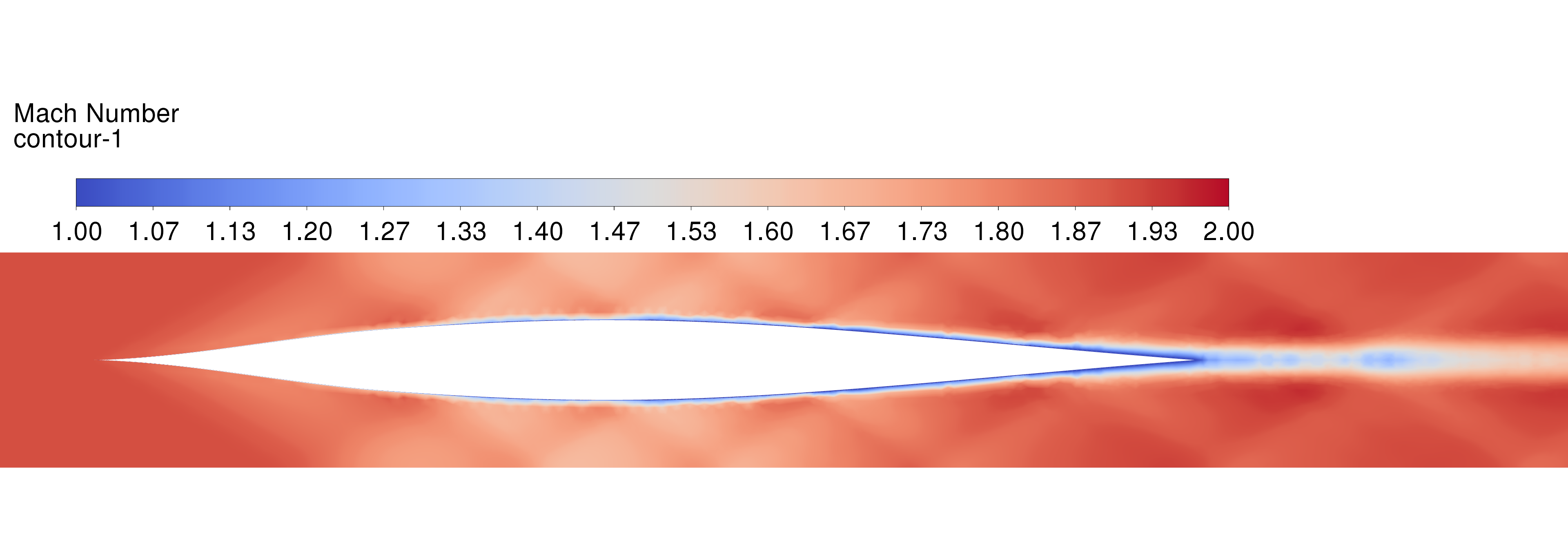}
\caption{Phase 2: $\beta=0.16$, $M_\infty=1.9$, $P_\infty=67583.775$ Pa.}
\label{fig:ss_s2_mach}
\end{subfigure}

\vspace{0.15cm}

\begin{subfigure}{\columnwidth}
\centering
\includegraphics[width=\linewidth]{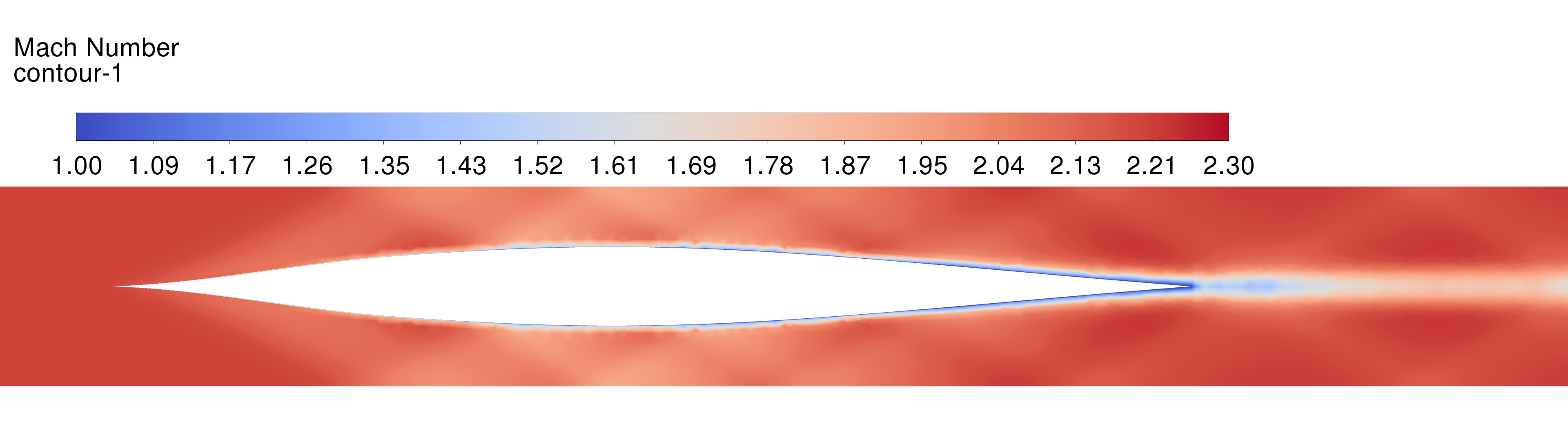}
\caption{Phase 3: $\beta=0.18$, $M_\infty=2.2$, $P_\infty=33842.55$ Pa.}
\label{fig:ss_s3_mach}
\end{subfigure}

\vspace{0.15cm}

\begin{subfigure}{\columnwidth}
\centering
\includegraphics[width=\linewidth]{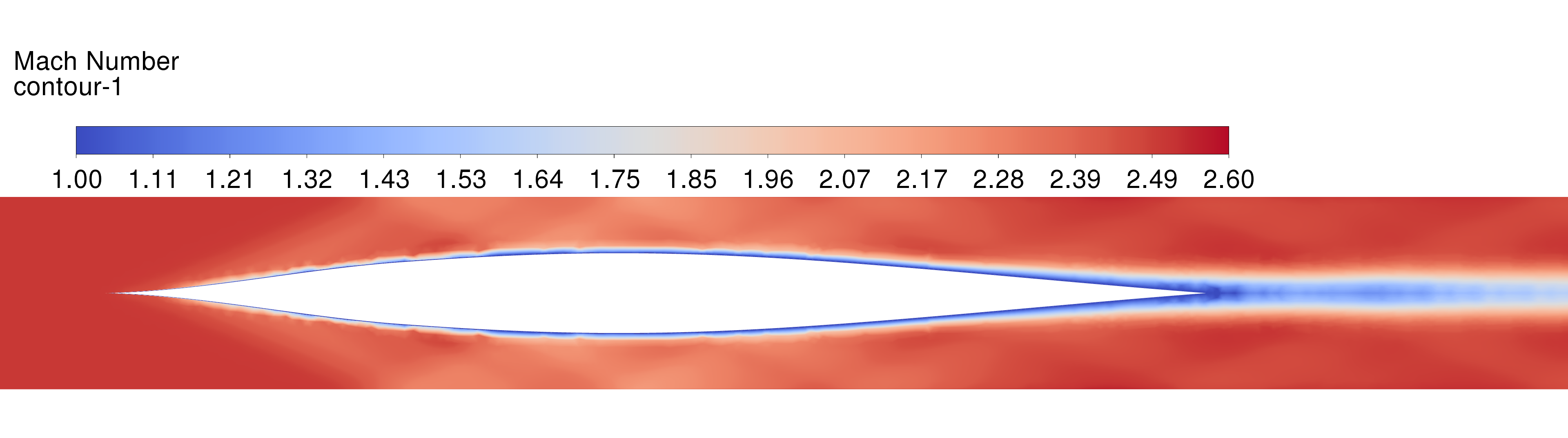}
\caption{Phase 4: $\beta=0.20$, $M_\infty=2.5$, $P_\infty=101.325$ Pa.}
\label{fig:ss_s4_mach}
\end{subfigure}

\vspace{0.15cm}

\begin{subfigure}{\columnwidth}
\centering
\includegraphics[width=\linewidth]{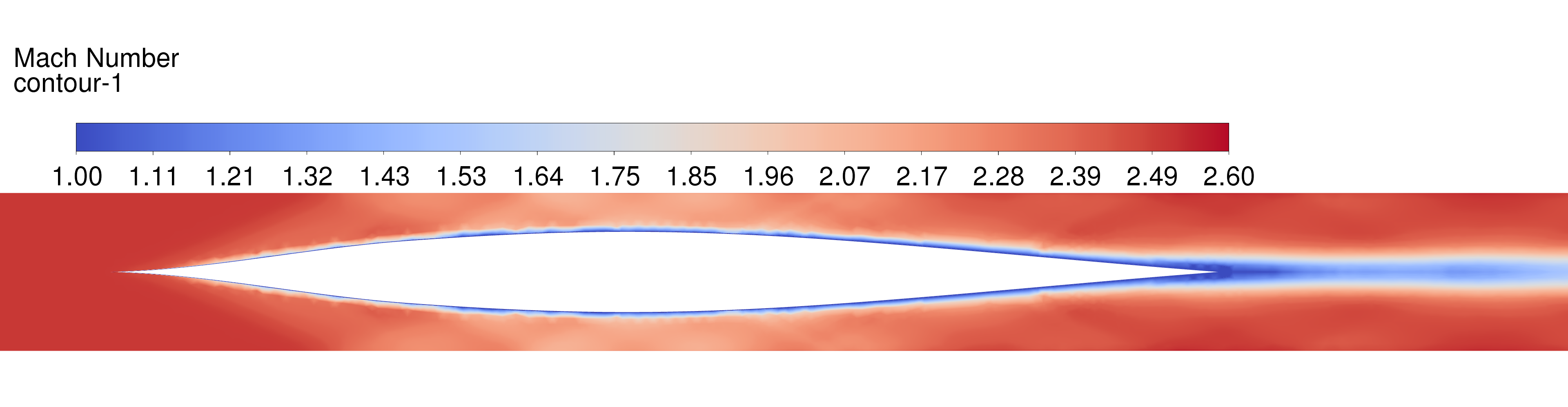}
\caption{Phase 5: $\beta=0.30$, $M_\infty=2.5$, $P_\infty=101.325$ Pa.}
\label{fig:ss_s5_mach}
\end{subfigure}

\vspace{0.15cm}

\begin{subfigure}{\columnwidth}
\centering
\includegraphics[width=\linewidth]{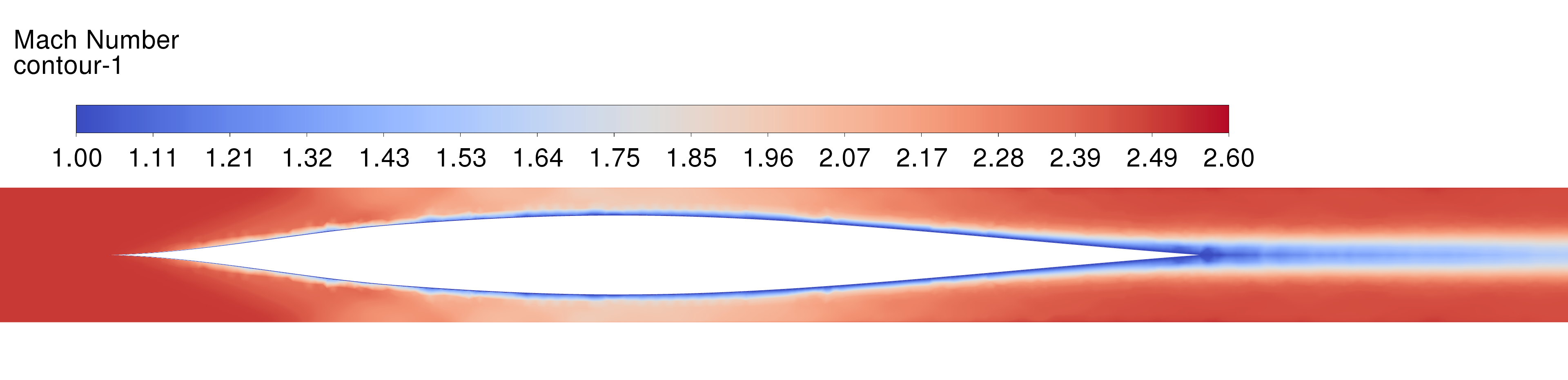}
\caption{Phase 6: $\beta=0.40$, $M_\infty=2.5$, $P_\infty=101.325$ Pa.}
\label{fig:ss_s6_mach}
\end{subfigure}

\vspace{0.15cm}

\begin{subfigure}{\columnwidth}
\centering
\includegraphics[width=\linewidth]{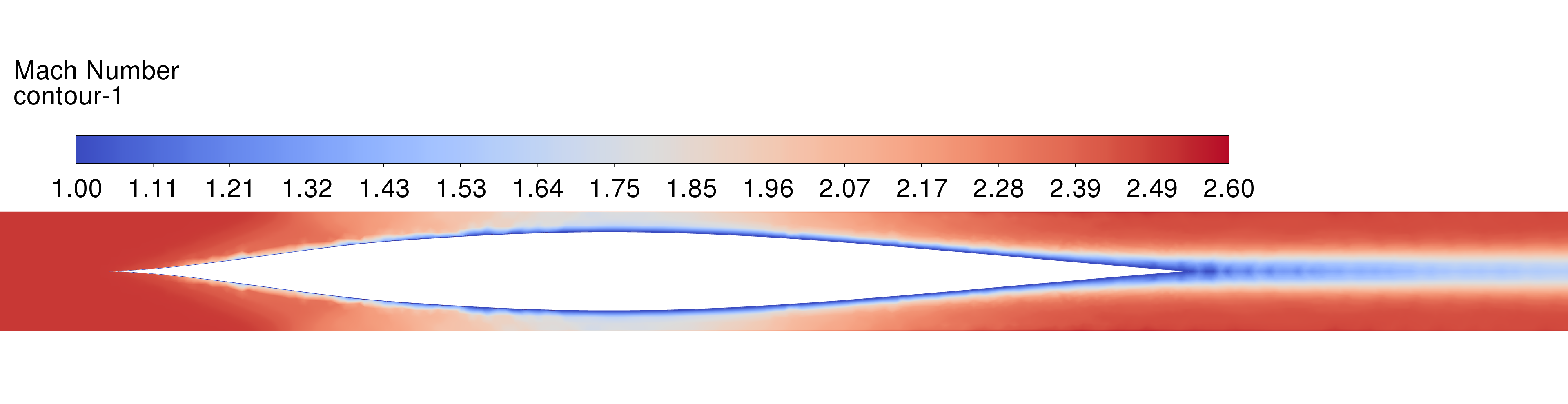}
\caption{Phase 7: $\beta=0.50$, $M_\infty=2.5$, $P_\infty=101.325$ Pa.}
\label{fig:ss_s7_mach}
\end{subfigure}

\caption{Mach number contours for the staged supersonic CD tube operating phases.}
\label{fig:ss_all_mach}
\end{figure}

\begin{figure}[!t]
\centering
\begin{subfigure}{\columnwidth}
\centering
\includegraphics[width=\linewidth]{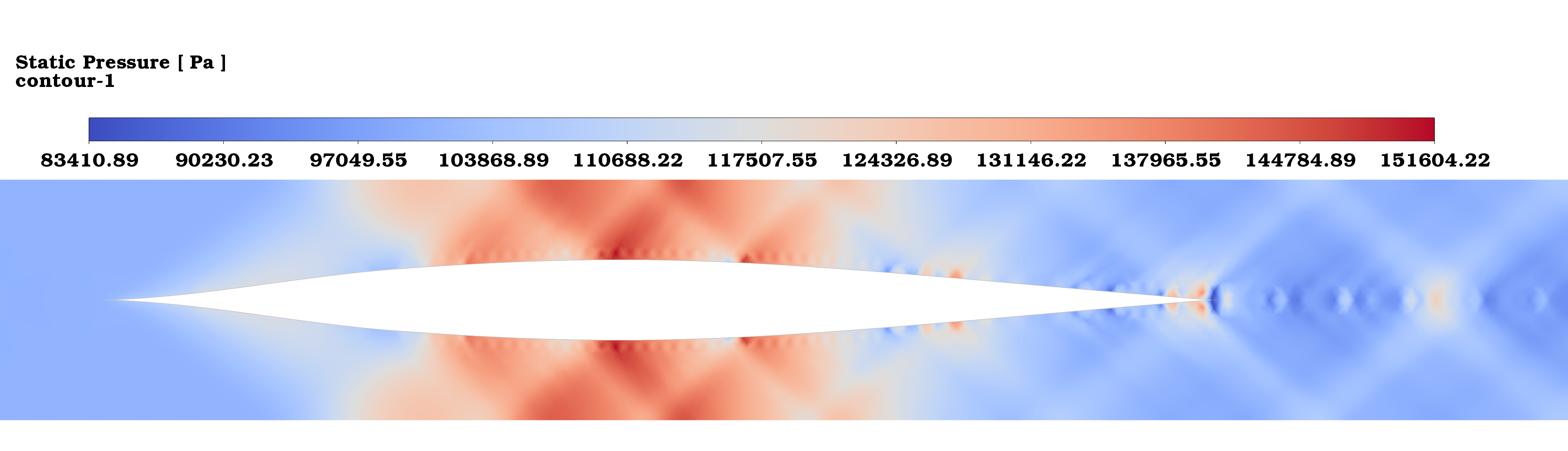}
\caption{Phase 1: $\beta=0.13$, $M_\infty=1.6$, $P_\infty=101325$ Pa.}
\label{fig:ss_s1_pres}
\end{subfigure}

\vspace{0.42cm}

\begin{subfigure}{\columnwidth}
\centering
\includegraphics[width=\linewidth]{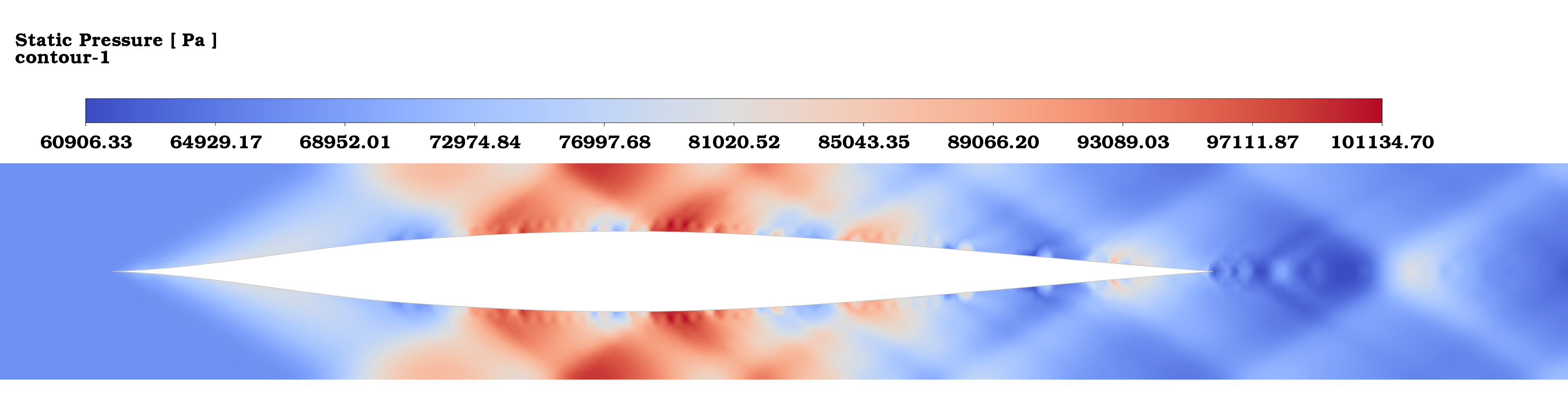}
\caption{Phase 2: $\beta=0.16$, $M_\infty=1.9$, $P_\infty=67583.775$ Pa.}
\label{fig:ss_s2_pres}
\end{subfigure}

\vspace{0.42cm}

\begin{subfigure}{\columnwidth}
\centering
\includegraphics[width=\linewidth]{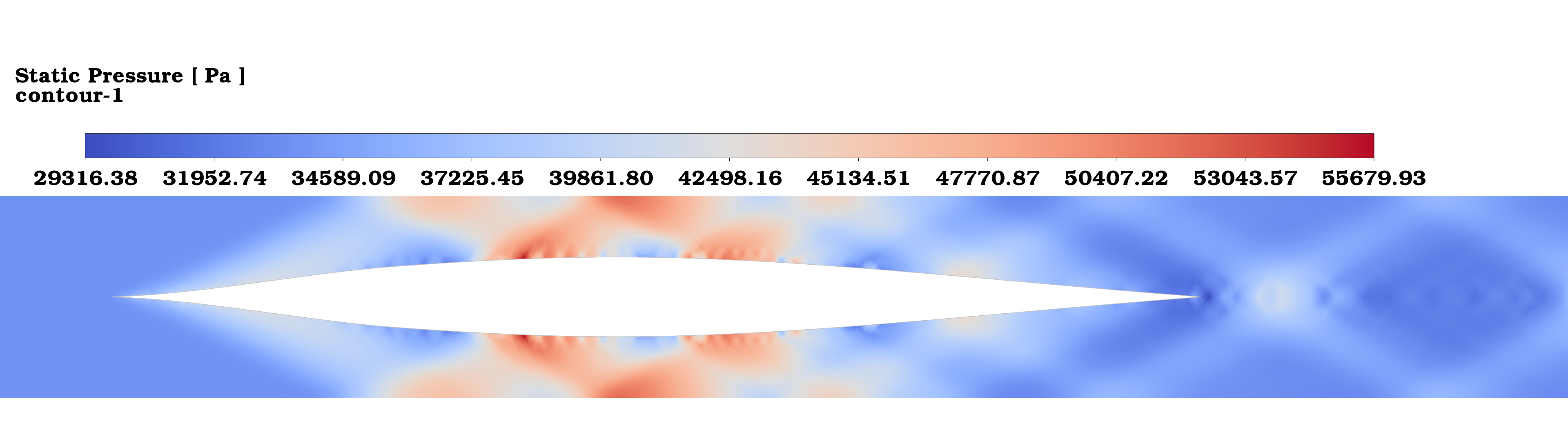}
\caption{Phase 3: $\beta=0.18$, $M_\infty=2.2$, $P_\infty=33842.55$ Pa.}
\label{fig:ss_s3_pres}
\end{subfigure}

\vspace{0.42cm}

\begin{subfigure}{\columnwidth}
\centering
\includegraphics[width=\linewidth]{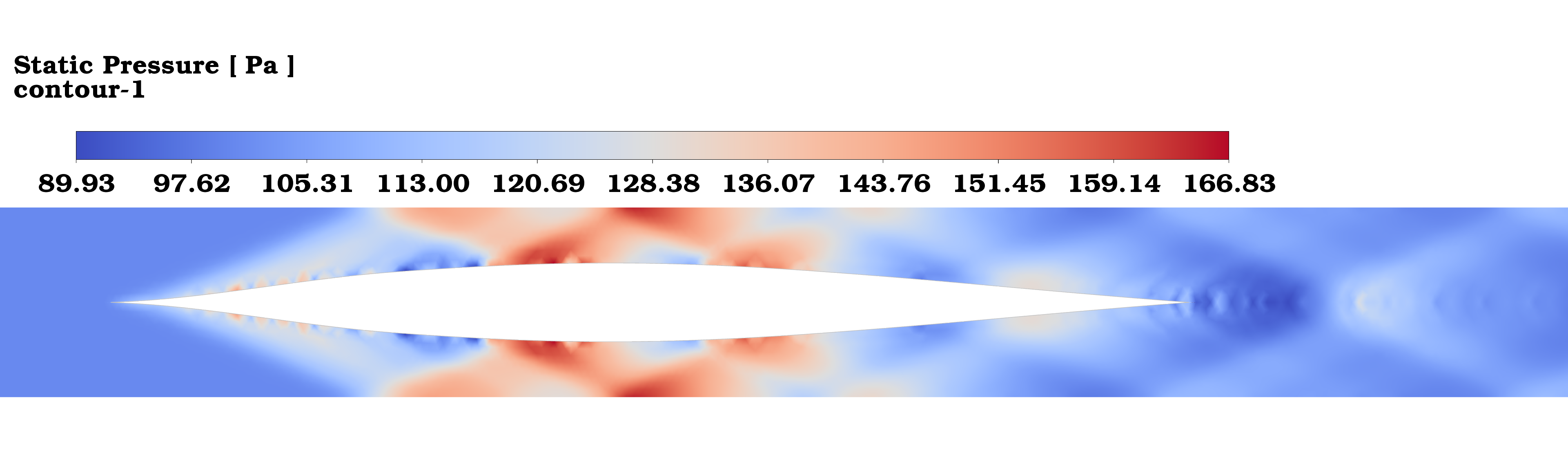}
\caption{Phase 4: $\beta=0.20$, $M_\infty=2.5$, $P_\infty=101.325$ Pa.}
\label{fig:ss_s4_pres}
\end{subfigure}

\vspace{0.42cm}

\begin{subfigure}{\columnwidth}
\centering
\includegraphics[width=\linewidth]{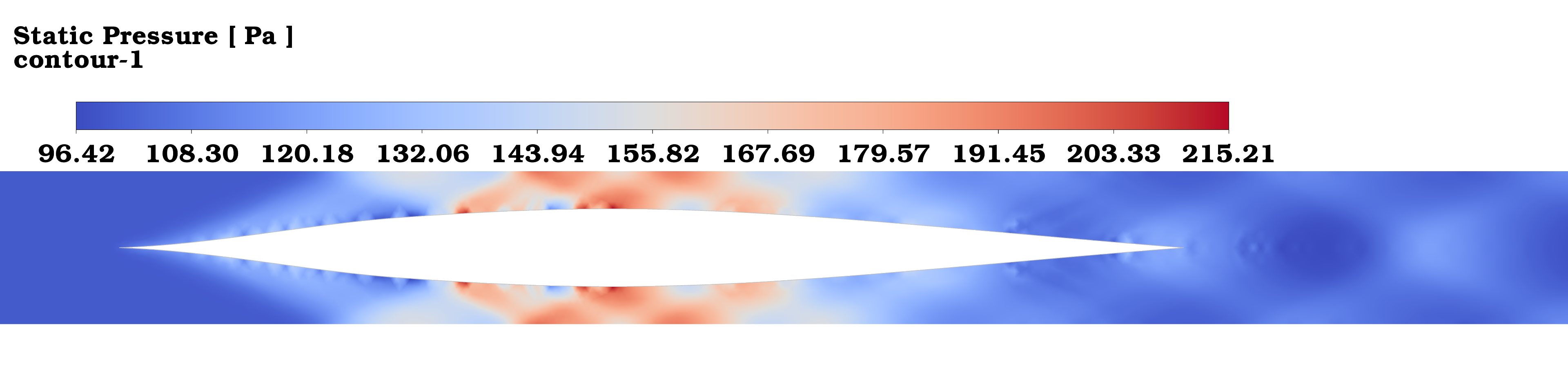}
\caption{Phase 5: $\beta=0.30$, $M_\infty=2.5$, $P_\infty=101.325$ Pa.}
\label{fig:ss_s5_pres}
\end{subfigure}

\vspace{0.42cm}

\begin{subfigure}{\columnwidth}
\centering
\includegraphics[width=\linewidth]{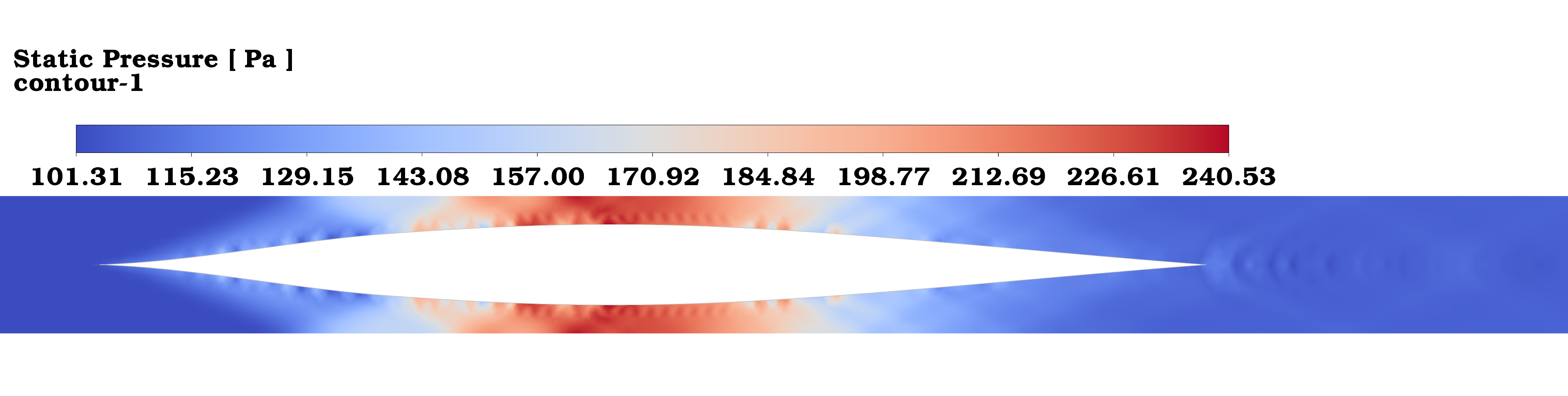}
\caption{Phase 6: $\beta=0.40$, $M_\infty=2.5$, $P_\infty=101.325$ Pa.}
\label{fig:ss_s6_pres}
\end{subfigure}

\vspace{0.42cm}

\begin{subfigure}{\columnwidth}
\centering
\includegraphics[width=\linewidth]{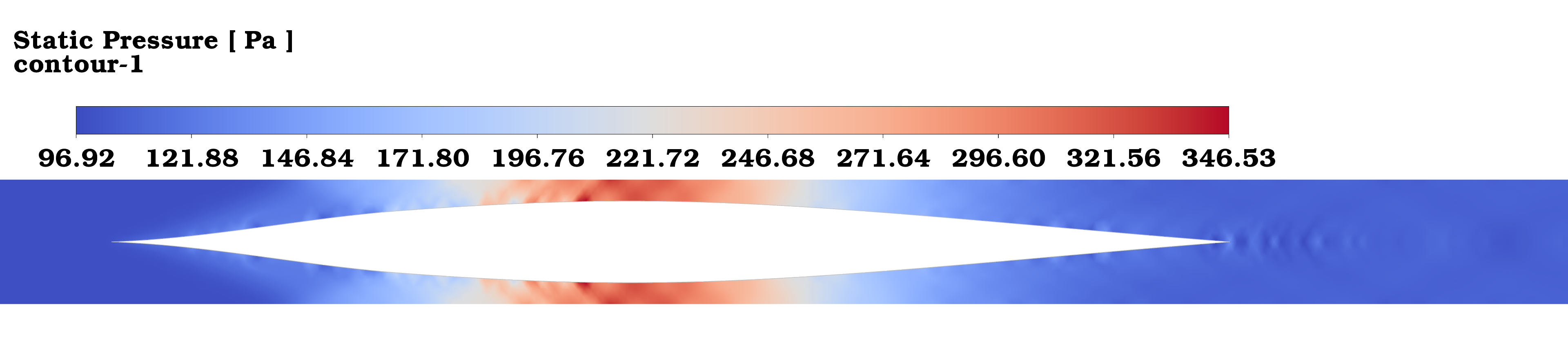}
\caption{Phase 7: $\beta=0.50$, $M_\infty=2.5$, $P_\infty=101.325$ Pa.}
\label{fig:ss_s7_pres}
\end{subfigure}

\caption{Static pressure contours for the staged supersonic CD tube operating phases.}
\label{fig:ss_all_pres}
\end{figure}

The Mach number and static pressure contours for the staged tube operating points are shown in Figures~\ref{fig:ss_all_mach} and~\ref{fig:ss_all_pres}. The simulations are arranged to represent the proposed operating sequence, beginning with low-blockage tube entry and progressing toward higher Mach number, lower pressure, and larger blockage ratio. Each operating point is initialized from the preceding solution using solution interpolation. This approach approximates a continuous operating path, although the simulations themselves are steady-state computations and should not be interpreted as a fully transient acceleration analysis.

At tube entry, the pod operates at $\beta = 0.10$, $M_\infty = 1.6$, and atmospheric pressure. The tube wall now confines the shock system and modifies the pressure field relative to the open-atmosphere baseline. However, the annular passage remains sufficiently wide that no sonic blockage is observed. The drag decreases to $22{,}479.1$\,N, primarily because the confined geometry modifies the wake and pressure-recovery behavior relative to the open baseline. This result also establishes the initial confined-flow condition for the staged sequence.

Increasing the blockage ratio to $\beta = 0.13$ at the same Mach number further modifies the shock reflections between the pod surface and the tube wall. The Mach and pressure contours in Figures~\ref{fig:ss_s1_mach} and~\ref{fig:ss_s1_pres} show reflected compression structures in the annular region, but no continuous sonic surface or normal-shock-induced blockage is observed. The total drag decreases to $19{,}998.1$\,N. This indicates that, at this stage, the improvement in pressure recovery outweighs the additional confinement penalty.

In Phase~2, the Mach number increases to $M_\infty = 1.9$, the blockage ratio increases to $\beta = 0.16$, and the pressure is reduced to $67{,}583.775$\,Pa. The higher Mach number strengthens the compressible-flow features, while the reduced pressure lowers the absolute aerodynamic loading. Figures~\ref{fig:ss_s2_mach} and~\ref{fig:ss_s2_pres} show a more complex pattern of incident and reflected shocks, together with pressure bands along the tube wall. The drag increases slightly to $21{,}893.7$\,N, suggesting that the stronger shock interactions and increased Mach number partly offset the benefit of pressure reduction at this intermediate stage.

In Phase~3, the operating point advances to $M_\infty = 2.2$, $\beta = 0.18$, and $P_\infty = 33{,}842.55$\,Pa. The contour fields in Figures~\ref{fig:ss_s3_mach} and~\ref{fig:ss_s3_pres} show stronger shock-reflection patterns and pressure nonuniformities within the annular passage. A supersonic wake structure appears downstream of the pod tail, indicating that the annular flow discharges into the downstream region with significant compressibility effects. The drag decreases to $16{,}384.1$\,N, mainly because the pressure reduction lowers the overall aerodynamic force scale. The wake structure is consistent with a supersonic annular discharge, but the choking assessment is based on the absence of a sustained sonic blockage or upstream normal-shock structure in the minimum-area region.

Phase~4 represents the transition to near-vacuum supersonic cruise, with $M_\infty = 2.5$, $\beta = 0.20$, and $P_\infty = 101.325$\,Pa. At this pressure, the absolute gas density is reduced by approximately three orders of magnitude relative to atmospheric conditions. Consequently, the pressure gradients and integrated aerodynamic loads are greatly reduced. The Mach contour in Figure~\ref{fig:ss_s4_mach} shows a comparatively uniform high-speed flow field with only weak compression features, while the pressure contour in Figure~\ref{fig:ss_s4_pres} spans approximately $89.93$--$166.83$\,Pa. The total drag decreases sharply to $123.6$\,N.

This large drag reduction should be interpreted carefully. It is not caused solely by pod-shape optimization or the CD tube concept; it is primarily a consequence of the staged reduction in ambient pressure, which reduces the density and therefore the aerodynamic force scale. Nevertheless, the result shows that the selected pressure--area path reaches the near-vacuum supersonic condition without producing a choked annular-flow structure within the assumptions of the steady axisymmetric CFD model.

\subsubsection{Near-Vacuum Blockage-Ratio Sweep}
\label{sec:blockage_sweep}

\begin{figure}[!b]
\centering
\includegraphics[width=0.89\columnwidth]{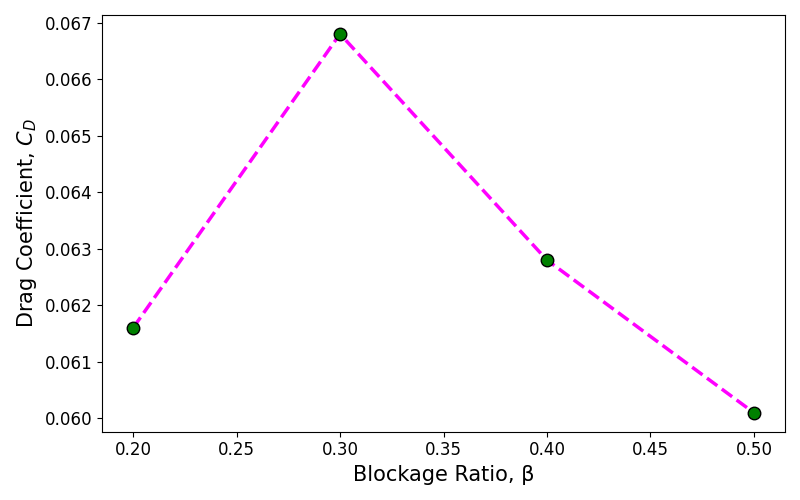}
\caption{Variation of drag coefficient $C_D$ with blockage ratio $\beta$ at $M_\infty = 2.5$ and $P_\infty = 101.325$\,Pa.}
\label{fig:cd_vs_beta}
\end{figure}

After reaching the near-vacuum cruise pressure, the blockage ratio is varied from $\beta = 0.20$ to $\beta = 0.50$ at fixed $M_\infty = 2.5$ and $P_\infty = 101.325$\,Pa. This sweep isolates the influence of tube-to-pod area ratio under low-density supersonic cruise conditions. The resulting variation of drag coefficient with blockage ratio is shown in Figure~\ref{fig:cd_vs_beta}.

Increasing $\beta$ from 0.20 to 0.30 narrows the annular passage and increases local acceleration near the pod surface. This raises wall shear stress and increases the drag coefficient to its maximum value in the sweep. The Mach contour for $\beta = 0.30$ in Figure~\ref{fig:ss_s5_mach} shows stronger annular acceleration than the $\beta = 0.20$ case, but no sustained sonic blockage is observed in the throat region.

For larger blockage ratios, $\beta = 0.40$ and $\beta = 0.50$, the drag coefficient decreases. The pressure contours in Figures~\ref{fig:ss_s6_pres} and~\ref{fig:ss_s7_pres} suggest improved pressure recovery near the aft portion of the pod as the tube wall more strongly influences the flow redirection around the tail. This pressure-recovery effect partly offsets the viscous penalty associated with the narrower annular passage. The minimum drag in the blockage sweep occurs at $\beta = 0.50$, where the total drag is $120.5$\,N and the corresponding drag coefficient is $C_D = 0.0601$.

This trend does not imply that increasing blockage ratio is generally beneficial for Hyperloop aerodynamics. Rather, it reflects the particular near-vacuum, high-Mach, axisymmetric operating conditions considered here. Under atmospheric or transonic conditions, increasing blockage ratio would generally increase choking risk and pressure drag. The present result therefore indicates that, once the flow is established in a low-density unchoked supersonic regime, higher blockage ratios may become aerodynamically acceptable within a limited range.

\subsubsection{Interpretation of Choking Behavior}
\label{sec:choking_interpretation}

Across the staged CD tube sequence, the CFD contours do not show the formation of a persistent sonic throat or a detached normal shock that would indicate a choked annular passage. The operating path therefore remains consistent with the unchoked branch described in Section~\ref{sec:theoretical_framework}. However, this conclusion should be interpreted within the limitations of the numerical approach. The present simulations are steady and axisymmetric, and the staged process is represented by a sequence of discrete operating points rather than a fully time-resolved acceleration through the tube.

A more complete demonstration of choking avoidance would require additional quantitative indicators, such as the maximum annular Mach number, minimum-area Mach number, mass-flow rate through the annular passage, and choking margin at each stage. Nevertheless, the present results provide preliminary evidence that staged pressure--area management can guide the pod from atmospheric supersonic entry toward near-vacuum supersonic cruise without triggering choking in the simulated operating path.

\begin{table}[htbp]
    \centering
    
    \label{tab:throat_mach}
    \begin{tabular}{lc}
        \toprule
        \textbf{Stage} & \textbf{Throat Mach Number} \\
        \midrule
        Phase 1 & 1.3970 \\
        Phase 2 & 1.6835 \\
        Phase 3 & 2.0029 \\
        Phase 4 & 2.2563 \\
        Phase 5 & 2.1559 \\
        Phase 6 & 1.9639 \\
        Phase 7 & 1.7655 \\
        \bottomrule
    \end{tabular}
    \caption{Annular throat Mach number at each staged operating point.}
\end{table}

\subsubsection{Overall Drag Reduction Across the Staged Sequence}
\label{sec:overall_drag_reduction}

The total drag decreases from $68{,}913.7$\,N in the atmospheric supersonic baseline to $120.5$\,N at the final near-vacuum cruise condition with $M_\infty = 2.5$ and $\beta = 0.50$. This corresponds to a reduction of more than two orders of magnitude. The reduction is dominated by the decrease in tube pressure from atmospheric conditions to near-vacuum conditions, which reduces the density and the dynamic pressure scale. The optimized pod geometry and blockage-ratio management further influence the distribution of pressure drag, viscous drag, and pressure recovery which can be observed in table \ref{tab:drag_phases}.

The results therefore support two main conclusions. First, the CFD--ML workflow can identify pod geometries whose surrogate-predicted drag reductions are consistent with CFD recomputation. Second, the staged CD tube concept provides a plausible aerodynamic pathway for reaching a high-Mach, near-vacuum cruise condition while avoiding choking in the idealized simulations considered here. These findings motivate further studies using transient three-dimensional CFD, refined choking-margin diagnostics, structural analysis of variable-area tube sections, airlock dynamics, propulsion-system integration, and full-system energy modelling.

\begin{table}[!t]
\centering
\small
\renewcommand{\arraystretch}{1.2}
\setlength{\tabcolsep}{5pt}

\begin{tabular}{p{4.5cm} c c c}
\hline
\textbf{Stage} & $\boldsymbol{\beta}$ &
\textbf{Pressure} &
\textbf{Viscous} \\
& & \textbf{drag [N]} &
\textbf{drag [N]} \\
\hline

Stage I: atmospheric baseline & ---  & 45,920.5 & 22,993.2 \\
Stage II: tube entry          & 0.10 &  1,841.7 & 20,637.4 \\
\hline

Phase 1 & 0.13 & $-$232.9 & 20,230.9 \\
Phase 2 & 0.16 &   374.7  & 21,519.1 \\
Phase 3 & 0.18 &   434.4  & 15,949.7 \\
Phase 4 & 0.20 &     0.8  &    122.7 \\
Phase 5 & 0.30 &  $-$12.0 &    145.9 \\
Phase 6 & 0.40 &  $-$24.4 &    150.3 \\
Phase 7 & 0.50 &  $-$34.4 &    154.9 \\
\hline
\end{tabular}

\caption{Pressure drag and viscous drag for each  
phase.}
\label{tab:drag_phases}

\end{table}

\FloatBarrier
\section{Transport Performance Benchmarking and Deployment Considerations}
\label{sec:transport_benchmarking}

The CFD results presented in Section~\ref{sec:results} quantify the aerodynamic drag associated with the optimized subsonic and supersonic ETT operating conditions. To place these values in a broader transport context, this section estimates a drag-only lower bound on specific energy consumption and compares it with representative values for established transport modes. The comparison is intended to provide an order-of-magnitude perspective rather than a complete system-level energy assessment.

\subsection{Drag-Based Specific Energy}
\label{subsec:specific_energy}

A convenient metric for comparing transport systems with different vehicle masses and passenger capacities is the energy expenditure per unit payload mass per unit distance, denoted here as $\varepsilon$ and expressed in \si{J\,kg^{-1}\,km^{-1}}. For the present ETT cases, the aerodynamic work required to overcome drag over a distance $\Delta x$ is

\begin{equation}
    E_{\mathrm{drag}} = D\,\Delta x,
\end{equation}

\noindent where $D$ is the total aerodynamic drag. The corresponding drag-based specific energy, normalized by the passenger payload mass $m_{\mathrm{pay}}$, is

\begin{equation}
    \varepsilon_{\mathrm{drag}}
    =
    \frac{D}{m_{\mathrm{pay}}}
    \quad
    \left[\si{J\,kg^{-1}\,m^{-1}}\right]
    =
    \frac{D \times 10^{3}}{m_{\mathrm{pay}}}
    \quad
    \left[\si{J\,kg^{-1}\,km^{-1}}\right].
    \label{eq:specific_energy}
\end{equation}

The payload mass is estimated using 24 passengers with a nominal mass of \SI{80}{kg} per passenger, giving

\begin{equation}
    m_{\mathrm{pay}} = 24 \times 80 = \SI{1920}{kg}.
\end{equation}

For the optimized supersonic near-vacuum cruise condition,
$M_\infty = 2.5$, $\beta = 0.50$, and $P_\infty = \SI{101.325}{Pa}$, the CFD-computed drag is $D = \SI{120.5}{N}$. The drag-based specific energy is therefore

\begin{equation}
    \varepsilon_{\mathrm{sup}}
    =
    \frac{120.5 \times 10^{3}}{1920}
    \approx
    \SI{63}{J\,kg^{-1}\,km^{-1}}.
\end{equation}

For the optimized subsonic near-vacuum cruise condition,
$M_\infty = 0.42$, $\beta = 0.36$, and $P_\infty = \SI{101.325}{Pa}$, the CFD-computed drag is $D = \SI{31.69}{N}$. This gives

\begin{equation}
    \varepsilon_{\mathrm{sub}}
    =
    \frac{31.69 \times 10^{3}}{1920}
    \approx
    \SI{16.5}{J\,kg^{-1}\,km^{-1}}.
\end{equation}

These values represent only the aerodynamic work required to overcome drag during steady cruise. They do not include vacuum generation and maintenance, propulsion losses, levitation power, auxiliary loads, acceleration and deceleration losses, station operation, or infrastructure-level energy costs. They should therefore be interpreted strictly as lower-bound aerodynamic indicators rather than estimates of total operational energy.

\subsection{Comparison with Conventional Transport Modes}
\label{subsec:energy_comparison}

Table~\ref{tab:energy_comparison} compares the drag-only ETT values with representative specific energy ranges for gasoline cars, commercial aviation, and high-speed rail. The conventional transport values are based on published fleet-average fuel consumption, emissions, occupancy, and traction-energy data \cite{icao2019,iea_uic2017,iea2023}. For cars, fuel use is converted to energy using gasoline density and lower heating value, with normalization by passenger payload mass. For aviation, passenger-kilometer emissions are converted to Jet-A fuel energy using standard emission factors. For high-speed rail, traction electricity use is converted directly from kWh to joules and normalized by passenger payload mass.

Because the ETT values in Table~\ref{tab:energy_comparison} include only aerodynamic drag during idealized cruise, whereas the conventional-mode values represent broader operational energy use, the table should not be interpreted as a direct full-system comparison. Instead, it highlights the magnitude of aerodynamic drag reduction achievable in a low-pressure tube environment.

\begin{table}[htbp]
\centering
\small
\renewcommand{\arraystretch}{1.25}

\begin{tabular*}{0.98\columnwidth}{@{\extracolsep{\fill}} p{5.8cm} c}
\hline
\textbf{Mode} & \textbf{$\varepsilon$ [J\,kg$^{-1}$\,km$^{-1}$]} \\
\hline
Gasoline car                          & 15{,}000--25{,}000 \\
Commercial aviation (A320 class)      & 12{,}000--18{,}000 \\
High-speed rail                       & 3{,}000--5{,}000 \\
\hline
ETT subsonic, drag-only lower bound   & $\approx 16.5$ \\
ETT supersonic, drag-only lower bound & $\approx 63$ \\
\hline
\end{tabular*}

\caption{Representative specific energy values for conventional transport modes compared with drag-only lower-bound estimates for the optimized ETT cases. The ETT values exclude system-level energy penalties and should not be interpreted as total operational energy consumption.}
\label{tab:energy_comparison}
\end{table}

The very low drag-based ETT values arise primarily from the near-vacuum operating pressure, which reduces gas density and hence aerodynamic force. This result supports the aerodynamic motivation for evacuated-tube operation. However, the practical energy performance of a full ETT system will depend on whether the additional energy costs associated with maintaining the low-pressure environment and operating the infrastructure remain small compared with the aerodynamic savings.

\subsection{System-Level Energy Penalties}
\label{subsec:system_penalties}

The lower-bound estimates above must be augmented by several system-level energy contributions before any full operational comparison can be made. The most important penalties include:

\begin{itemize}
    \item \textbf{Vacuum generation and maintenance:} Distributed pumping stations are required to initially evacuate the tube and continuously compensate for leakage through seals, joints, stations, and airlock interfaces. The magnitude of this penalty depends strongly on route length, tube diameter, leakage rate, pump efficiency, and operating pressure.

    \item \textbf{Magnetic levitation and guidance:} Continuous levitation, lateral stabilization, and guideway control require electrical power. Additional losses may arise from eddy currents, control-system actuation, and high-speed guideway interactions.

    \item \textbf{Propulsion efficiency:} Linear motors, inverters, power electronics, and transmission systems introduce conversion losses. These losses must be included when converting aerodynamic drag power into grid-level energy demand.

    \item \textbf{Acceleration and deceleration:} The kinetic energy required to accelerate the pod scales with total vehicle mass and the square of the cruise velocity. A fraction of this energy may be recovered through regenerative braking, but recovery efficiency, storage limitations, and scheduling constraints will determine the net penalty.

    \item \textbf{Cabin pressurization and thermal management:} The passenger cabin must remain pressurized and thermally controlled while operating inside a near-vacuum environment. These auxiliary loads become particularly important for long routes and high-frequency operation.

    \item \textbf{Station and airlock operation:} Passenger boarding, pod dispatch, pressure equalization, and staged airlock operation introduce additional energy and time penalties that are not represented in cruise-only aerodynamic calculations.
\end{itemize}

Depending on route length, leakage rate, duty cycle, and technology maturity, these system-level contributions may increase the total energy consumption by several times relative to the drag-only estimate. A rigorous full-system assessment would require coupled modelling of aerodynamics, propulsion, levitation, vacuum infrastructure, airlock operation, thermal loads, traffic frequency, and regenerative braking.

\subsection{Representative Intercity Travel-Time Estimates}
\label{subsec:travel_times}

The supersonic ETT concept is also evaluated in terms of indicative travel time. At a cruise Mach number of $M = 2.5$, the nominal cruise speed is approximately \SI{3125}{km/h}. Assuming a fixed 20-minute allowance for staged acceleration, deceleration, and pressure-transition operations, and assuming that approximately 300\,km of the route is associated with non-cruise operation, the total travel time for a route of length $D$ can be approximated as

\begin{equation}
    t_{\mathrm{total}}
    \approx
    20~\text{min}
    +
    \frac{D - 300}{3125}
    \times 60~\text{min}.
    \label{eq:travel_time}
\end{equation}

This expression is intended only as a first-order estimate. It assumes that the route is long enough to include a meaningful cruise segment and does not account for station dwell time, route curvature, speed restrictions, safety spacing, network congestion, or passenger-access time. Representative travel-time estimates are shown in Table~\ref{tab:ett_routes}.
\begin{table}[!t]
\centering
\small
\renewcommand{\arraystretch}{1.2}
\setlength{\tabcolsep}{6pt}

\begin{tabular}{p{2.7cm} c c c}
\hline
\textbf{Route} &
\textbf{Dist.} &
\textbf{ETT} &
\textbf{Flight} \\
&
\textbf{[km]} &
\textbf{estimate} &
\textbf{time} \\
\hline

\multicolumn{4}{l}{\textit{India}} \\
Mumbai--Delhi   & 1153 & $\sim$36 min & 2 h 10 min \\
Mumbai--Kolkata & 1668 & $\sim$46 min & 2 h 20 min \\
Delhi--Chennai  & 1754 & $\sim$48 min & 2 h 45 min \\

\hline
\multicolumn{4}{l}{\textit{United States}} \\
New York--Chicago & 1149 & $\sim$36 min & 2 h 25 min \\
Miami--Seattle    & 4395 & $\sim$99 min & 6 h 15 min \\

\hline
\multicolumn{4}{l}{\textit{Europe}} \\
Madrid--Berlin & 1869 & $\sim$50 min & 3 h 10 min \\
Lisbon--Warsaw & 2754 & $\sim$67 min & 4 h 00 min \\

\hline
\end{tabular}

\caption{First-order travel-time estimates for representative routes assuming $M = 2.5$ cruise and a fixed 20-minute allowance for acceleration, deceleration, and pressure-transition operations. Estimates exclude station dwell time, route curvature, operational speed limits, and passenger-access time.}
\label{tab:ett_routes}
\end{table}

The estimates indicate the potential time advantage of supersonic ETT on long intercity routes. However, the practical benefit depends strongly on total door-to-door travel time, station placement, route alignment, acceleration limits for passenger comfort, and operational constraints. For shorter routes, the non-cruise fraction becomes large, reducing the advantage of very high cruise speed.

\subsection{Outstanding Engineering Challenges}
\label{subsec:deployment_challenges}

The present study focuses on aerodynamic feasibility and does not resolve the engineering requirements for deployment. Several major challenges must be addressed before a supersonic ETT system could be considered operationally viable:

\begin{itemize}
    \item \textbf{Tube structural integrity and alignment:} Long evacuated tubes must withstand external atmospheric pressure, thermal expansion, seismic loading, ground settlement, and structural vibration while maintaining tight geometric tolerances compatible with high-speed operation.

    \item \textbf{Vacuum sealing and airlock reliability:} Maintaining low pressure over long distances requires robust seals, distributed pumping, reliable airlock operation, and fault-tolerant pressure-management systems.

    \item \textbf{Emergency evacuation and rescue:} A stationary pod inside a near-vacuum tube cannot be evacuated using conventional rail or aviation procedures. Emergency refuge spacing, re-pressurization protocols, rescue access, and passenger life-support duration must be defined and validated.

    \item \textbf{Thermal and aeroelastic effects:} High-speed operation in a confined tube may introduce thermal gradients, structural vibration, acoustic loading, and fluid--structure interaction effects that are not captured in the present steady axisymmetric simulations.

    \item \textbf{Propulsion, levitation, and control integration:} The aerodynamic design must be integrated with linear propulsion, levitation, braking, guidance, and stability-control systems, all of which affect vehicle mass, energy use, safety margins, and operational cost.

    \item \textbf{Certification and regulation:} ETT does not fit cleanly within existing aviation, rail, or maritime regulatory frameworks. A dedicated safety, certification, and operations framework would be required before commercial deployment.
\end{itemize}

These considerations emphasize that the low aerodynamic drag predicted in this study is a necessary but not sufficient condition for practical supersonic ETT. The results provide an aerodynamic basis for further investigation, but full feasibility depends on coupled system-level analysis across infrastructure, propulsion, safety, energy, and operations.

\section{Conclusion}
\label{sec:conclusion}

This study developed a CFD--machine learning framework for aerodynamic optimization of Hyperloop pod geometries and examined a staged converging--diverging (CD) tube concept for supersonic evacuated tube transport. The work focused on reducing aerodynamic drag while mitigating flow choking in the confined annular passage between the pod and tube wall. Within the assumptions of steady two-dimensional axisymmetric CFD, the results provide a concept-level aerodynamic assessment of combined pod-shape optimization and staged pressure--area management.

The CFD--ML framework enabled efficient exploration of the pod design space. Pod geometries were parameterized using composite cubic B\'{e}zier curves and evaluated using validated ANSYS Fluent simulations. The resulting CFD dataset was used to train surrogate models for drag prediction. XGBoost achieved the best overall performance, with a test $R^2$ of 0.9524. For the optimized subsonic geometry, the surrogate-predicted drag differed from CFD recomputation by only 1.58\%, confirming that the surrogate could reliably guide optimization within the sampled design bounds.

For the subsonic case at $M_\infty = 0.42$ and $P_\infty = 101.325$\,Pa, the optimized pod produced a CFD-computed drag of 31.69\,N. The flow remained fully subsonic, with smooth acceleration through the annular passage and partial pressure recovery along the tail. The wall shear stress remained positive over the full pod length, indicating no boundary-layer separation for the optimized geometry.

For supersonic operation, the staged CD tube concept was used to guide the pod from atmospheric entry at $M_\infty = 1.6$ to near-vacuum cruise at $M_\infty = 2.5$ and $\beta = 0.50$. In the steady axisymmetric simulations, the selected pressure--area path did not show a persistent sonic throat or detached normal shock indicative of annular-flow choking. The total drag decreased from 68{,}913.7\,N in the atmospheric supersonic baseline to 120.5\,N at the final near-vacuum cruise condition, primarily due to the large reduction in operating pressure. A blockage-ratio sweep at near-vacuum cruise showed the lowest drag coefficient in the investigated range at $\beta = 0.50$, with $C_D = 0.0601$.

The drag-only energy estimates indicate the aerodynamic advantage of low-pressure ETT operation, but they represent strict lower bounds and do not include vacuum maintenance, propulsion losses, levitation power, airlock operation, acceleration/deceleration losses, or station-level energy demands. Thus, the present results establish aerodynamic plausibility, not full system feasibility.

Overall, the study demonstrates that CFD-driven surrogate optimization can identify low-drag Hyperloop pod geometries and that staged pressure--area management is a promising concept for mitigating choking in supersonic ETT operation. Future work should include transient three-dimensional CFD, quantitative choking-margin analysis, refined grid-convergence studies, airlock and vacuum-system modeling, propulsion--levitation integration, and full-system energy and cost assessment.

\section*{Declaration of Competing Interest}
The authors declare that they have no known competing financial 
interests or personal relationships that could have appeared to 
influence the work reported in this paper. All research was 
conducted independently, without any commercial, financial, or 
institutional influence that might have biased the outcomes or 
interpretation of the results presented herein.

\section*{Data Availability }
The data supporting the findings of this study, including 
computational fluid dynamics (CFD) simulation parameters, 
boundary conditions, mesh configurations, and post-processed 
results, are available from the corresponding author upon 
reasonable request.

\section*{Declaration of Generative AI and AI-Assisted Technologies in the Manuscript Preparation Process}

During the preparation of this work the author(s) used Claude (Anthropic) in order to improve the language and readability of the manuscript. After using this tool/service, the author(s) reviewed and edited the content as needed and take(s) full responsibility for the content of the published article.

\FloatBarrier
\nocite{*}
\bibliographystyle{unsrt}
\bibliography{references}

\end{document}